\documentclass[twocolumn, prx, superscriptaddress,notitlepage]{revtex4-1}
\usepackage{graphicx}
\usepackage[usenames,dvipsnames]{xcolor}
\usepackage[utf8]{inputenc}
\usepackage[T1]{fontenc}
\usepackage{amsfonts}
\usepackage{gensymb}
\usepackage[normalem]{ulem}
\usepackage{comment}
\usepackage[colorlinks, linkcolor=blue,anchorcolor=blue,citecolor=blue,urlcolor=blue]{hyperref}
\usepackage{amssymb}
\usepackage{pifont}
\usepackage{physics}
\usepackage{natbib}
\usepackage{array}
\usepackage{multirow}

\begin{document}
\title{Exploring the mechanisms of transverse relaxation of copper(II)-phthalocyanine spin qubits}

\author{Boning Li}\thanks{These authors contributed equally.}
    \affiliation{Department of Physics, Massachusetts Institute of Technology, Cambridge, MA 02139, USA}
    \affiliation{Research Laboratory of Electronics, Massachusetts Institute of Technology, Cambridge, MA 02139, USA}
    
\author{Yifan Quan}\thanks{These authors contributed equally.}
    \affiliation{Department of Chemistry and Francis Bitter Magnet Laboratory, Massachusetts Institute of Technology, Cambridge, MA 02139, USA}
\author{Xufan Li}\thanks{These authors contributed equally.}
   \affiliation{Honda Research Institute USA, Inc., San Jose, CA 95134, USA}
\author{Guoqing Wang}
    \affiliation{Research Laboratory of Electronics, Massachusetts Institute of Technology, Cambridge, MA 02139, USA}
    \affiliation{Department of Physics, Massachusetts Institute of Technology, Cambridge, MA 02139, USA}
\author{Robert G Griffin}
    \affiliation{Department of Chemistry and Francis Bitter Magnet Laboratory, Massachusetts Institute of Technology, Cambridge, MA 02139, USA}
    \author{Avetik R Harutyunyan}\email{aharutyunyan@honda-ri.com}
    \affiliation{Honda Research Institute USA, Inc., San Jose, CA 95134, USA}
    \affiliation{Department of Nuclear Science and Engineering, Massachusetts Institute of Technology, Cambridge, MA 02139, USA}
\author{Paola Cappellaro}\email[]{pcappell@mit.edu}
    \affiliation{Department of Physics, Massachusetts Institute of Technology, Cambridge, MA 02139, USA}
    \affiliation{Research Laboratory of Electronics, Massachusetts Institute of Technology, Cambridge, MA 02139, USA}
    \affiliation{Department of Nuclear Science and Engineering, Massachusetts Institute of Technology, Cambridge, MA 02139, USA}

\begin{abstract}

Molecular spin qubits are promising candidates for quantum technologies, but their performance is limited by decoherence arising from diverse mechanisms. The complexity of the environment makes it challenging to identify the main source of noise and target it for mitigation. 
Here we present a systematic experimental and theoretical framework for analyzing the mechanisms of transverse relaxation in copper(II) phthalocyanine (CuPc) diluted into diamagnetic phthalocyanine hosts.
Using pulsed EPR spectroscopy together with first-principles cluster correlation expansion simulations, we quantitatively separate the contributions from hyperfine-coupled nuclear spins, spin--lattice relaxation, and electron--electron dipolar interactions.
Our detailed modeling  shows that both strongly and weakly coupled nuclei contribute negligibly to $T_2$, while longitudinal dipolar interactions with electronic spins, through instantaneous and spectral diffusion, constitute the main decoherence channel even at moderate spin densities.
This conclusion is validated by direct comparison between simulated spin-echo dynamics and experimental data.
By providing a robust modeling and experimental approach, our work  identifies favorable values of the electron spin density for quantum applications, and provides a transferable methodology for predicting ensemble coherence times.
These insights will guide the design and optimization of molecular spin qubits for scalable quantum devices.

\end{abstract}

\maketitle

\section{Introduction}

Quantum coherence in spin systems underpins the operation of quantum technologies ranging from quantum information processing and sensing to hybrid quantum materials~\cite{nielsen2002quantum,wu2022enhanced,takahashi2011decoherence,godfrin2017operating}.
Identifying and mitigating the sources of decoherence in solid-state spin ensembles is therefore a central challenge.
Among different platforms, molecular spin qubits are attractive because of their chemical tunability, structural reproducibility, and potential for scalable integration~\cite{bertaina2008quantum, warner2013potential, atzori2016room, atzori2019second}.
In particular, transition-metal phthalocyanines (Pc) provide a versatile family of candidate systems that can be synthesized in both magnetic and diamagnetic forms, enabling precise control of spin concentration and host environment~\cite{de2013voyage,de2009functional,warner2013potential}.

In such molecular qubits, the electron spin coherence time ($T_2$) is limited by interactions with the surrounding environment.
Historically, three decoherence channels have been considered: (i) hyperfine interactions with nearby nuclear spins, (ii) electron--electron dipolar couplings that generate instantaneous and spectral diffusion, and (iii) spin--lattice relaxation processes.
Prior studies of molecular magnets and organic radicals have examined the influence of ligands~\cite{mirzoyan2021deconvolving,wedge2012chemical} and solvents~\cite{graham2017probing,gustin2023mapping}. However, existing spin-bath models ~\cite{stamp2004coherence, warner2013potential} relied on simplified models that lacked quantitative validation against experimental data, and key parameters such as electron spin density were not accurately determined.
As a result, the relative importance of these different mechanisms has remained unsettled.

Beyond their role as model spin qubits, phthalocyanines and related molecular magnets occupy a unique position at the intersection of physics, chemistry, and materials science.
Their long spin lifetimes and tunable electronic structure have already been harnessed in spintronics and molecular data storage, where coherence directly governs information retention and manipulation~\cite{bogani2008molecular}, and in quantum information and sensing, where it serves as the essential figure of merit for reliable qubit operation~\cite{wasielewski2020exploiting}.
At the same time, recent studies highlight that spin alignment and coherence can influence catalytic reactivity and magnetic-field--driven chemistry, offering new routes to spin-selective catalysis~\cite{cao2023spin,bordet2025magnetically}.
Furthermore, cobalt phthalocyanine--based magnets have even been explored for biomedical applications, where their magnetic and spin properties could be directly leveraged in medical technologies~\cite{Harutyunyan1999JMMM}.
The ability to quantify and predict decoherence, as we demonstrate here for CuPc, thus holds implications far beyond qubits, informing the design of functional molecular materials for spintronics, catalysis, medicine, and energy technologies~\cite{de2013voyage,de2009functional}.

Here, we study decoherence by developing a systematic framework that combines quantum noise spectroscopy-inspired~\cite{suter2016colloquium,cywinski2014dynamical,yan2013rotating} pulsed EPR experiments with first-principles cluster correlation expansion (CCE) simulations~\cite{yang2008quantum,onizhuk2021pycce} to quantitatively dissect the contributions of nuclear, dipolar, and lattice environments in copper(II)--phthalocyanine (CuPc) diluted into diamagnetic phthalocyanine matrices.
This approach allows us to move beyond phenomenological descriptions and to establish experimentally validated microscopic parameters.

By explicitly separating strongly hyperfine-coupled nuclei (Cu, N) from weakly coupled ones (e.g.\ H), we show that both play only a negligible role in limiting $T_2$,  contrary to the prevailing assumption in earlier literature.
Instead, we find that electron--electron dipolar interactions, through both instantaneous and spectral diffusion, constitute the dominant decoherence pathway even at moderate spin densities.
This conclusion is supported by direct comparisons between simulated spin-echo dynamics and measured EPR decays, which show quantitative agreement across a range of spin concentrations.
Importantly, our results correct earlier models and provide a reliable route to experimentally estimate  electron spin density.

This work not only confirms that dipolar interactions are important, an observation consistent with trends in other spin systems, but establishes a rigorous, predictive methodology for disentangling decoherence channels.
This framework can be readily transferred to other molecular spin qubits, offering a pathway to rationally design host--guest systems with improved coherence.
Moreover, the ability to quantitatively evaluate competing decoherence mechanisms is relevant to broader contexts in quantum information science, including defect spins in solids, molecular magnets, and organic radical systems.

In the following, we describe the synthesis and characterization of CuPc diluted in diamagnetic XPc hosts, present spin-echo and spin-locking measurements of coherence times, and develop CCE simulations that incorporate nuclear and electronic spin baths.
We then compare theory and experiment to identify the dominant decoherence channels and discuss the implications for molecular qubit design.

\section{Experimental System and Overview of Decoherence Mechanisms \label{section:mechanics}}

\begin{figure}[htbp]
\includegraphics[width=0.46\textwidth]{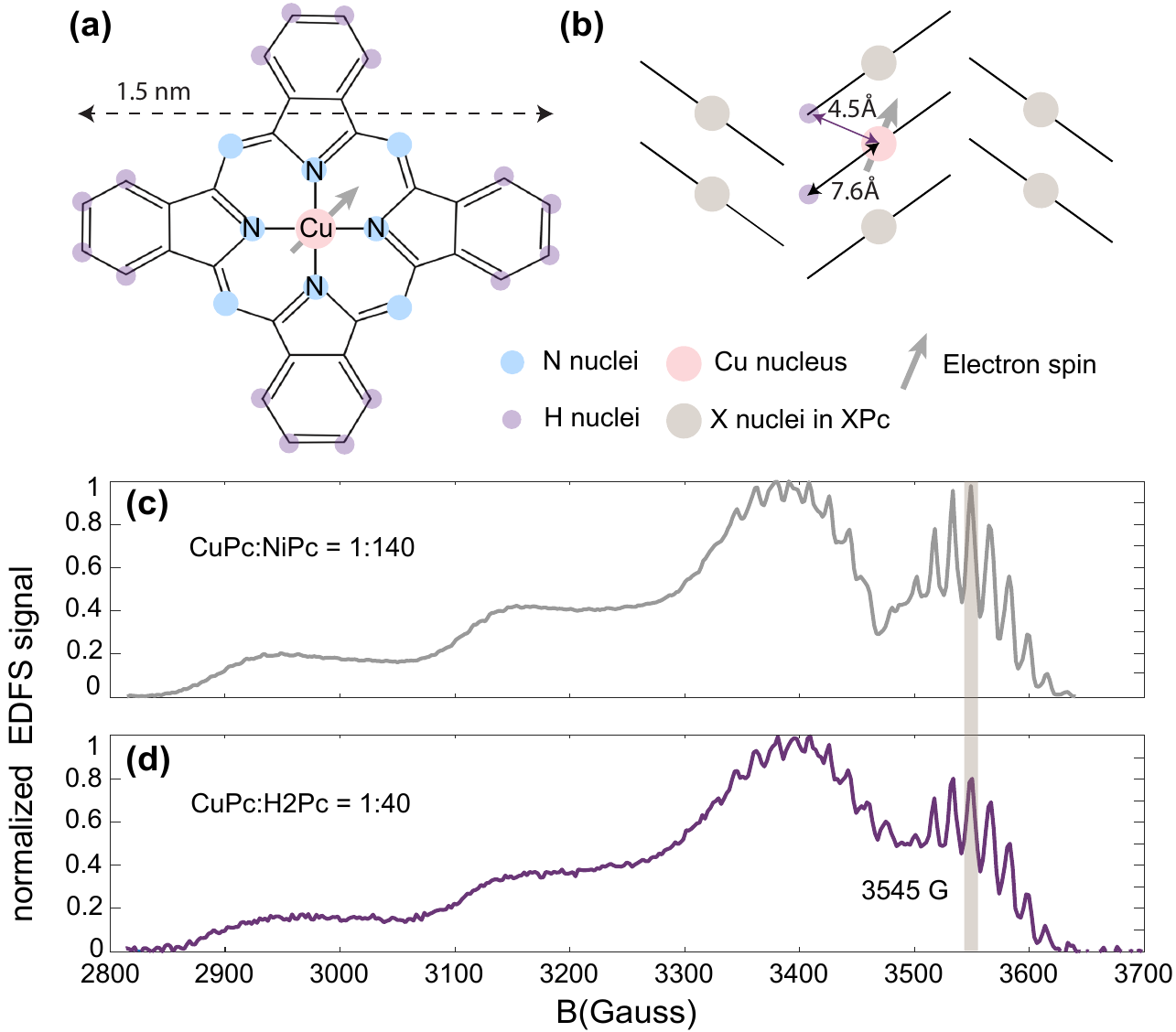}
 \caption{\textbf{Structural and spectral characterization of CuPc molecular systems.} 
    (a) Structure of a copper phthalocyanine (CuPc) molecule. 
    (b) Lattice structure of $\beta$-phase CuPc:XPc crystal, where XPc denotes diamagnetic phthalocyanine (e.g., NiPc or H$_2$Pc).  We indicate the distances of the Cu e- spin to the nearest intra- and inter-molecule protons, the second being the smallest. 
    (c,d) Echo-Detected Field Sweep (EDFS) measurement results for CuPc:NiPc and CuPc:H$_2$Pc samples, respectively, taken at a fixed resonance frequency of 9.72~GHz.  One of the EPR transitions ($\sim$3545~G) split by hyperfine couplings is highlighted. 
    }
\label{fig:1}
\end{figure}

Copper(II)-phthalocyanine (CuPc) is a planar molecule that hosts an $S = 1/2$ electron spin localized on the Cu$^{2+}$ ion. The spin exhibits $g$-anisotropy ($g_\perp\approx 2.04$,  $g_\parallel\approx2.16$~\cite{finazzo2006matrix}) and hyperfine interactions with  the copper  and nearby nitrogen nuclei. Copper has two naturally abundant isotopes, $^{63}\mathrm{Cu}$ (69.15\%) and $^{65}\mathrm{Cu}$ (30.85\%), both with nuclear spin $I = \tfrac{3}{2}$. 
The hyperfine coupling constants for $^{63}\mathrm{Cu}$ are $A_{xx}^{\mathrm{Cu}} = A_{yy}^{\mathrm{Cu}} = -83$~MHz and $A_{zz}^{\mathrm{Cu}} = -648$~MHz and the quadrupolar interaction $Q\approx3$~MHz~\cite{finazzo2006matrix}. The hyperfine interaction strength for $^{65}\mathrm{Cu}$ is rescaled by its gyromagnetic ratio, ${\gamma_{^{65}\mathrm{Cu}}}/{\gamma_{^{63}\mathrm{Cu}}} = 1.07$~\cite{stone2019table}. 
The Cu$^{2+}$ electron spin also couples strongly to the four nearest $^{14}\mathrm{N}$ nuclear spins ($I = 1$), with hyperfine constants $A_{xx}^{\mathrm{N}} = 57$~MHz and $A_{yy}^{\mathrm{N}} = A_{zz}^{\mathrm{N}} = 45$~MHz~\cite{finazzo2006matrix}. The molecular structure of CuPc is shown in Fig.~\ref{fig:1}(a). The four non-bonded nitrogen nuclei and the hydrogen nuclear spins, whose couplings to the Cu nuclear spin are on the order of hundreds of kHz, are neglected for the moment. Their contributions to decoherence are discussed in Section~\ref{section:nuclear}.  

To suppress inter-electron spin interactions that limit coherence ~\cite{li25cupcnvpaper,warner2013potential}, CuPc is diluted into diamagnetic host matrices such as NiPc or H$_2$Pc, forming CuPc:XPc crystals. These hosts co-crystallize with CuPc in the $\beta$-phase, forming a herringbone-stacked structure (Fig.~\ref{fig:1}(b)) that ensures uniform dispersion and preserves the molecular environment~\cite{hammond1996x,luis2014molecular}.  

In the applied magnetic field ($B_0\approx3545$~G) the single-molecule Hamiltonian is
\begin{align}
    {\cal H}_{\textrm{CuPc}} =& \beta_e B_0\vec{z'} \cdot \mathbf{g} \cdot \vec{S} + \sum_n \vec{S} \cdot \textbf{A}_n \cdot \vec{I}_n \\
    &+ B_0\vec{z'}\cdot \sum_n \gamma_n \vec{I}_n+\sum_n\mathbf{Q}_n\cdot \vec I_{n}\\
    &+ \sum_{n_i,n_j}\vec{I}_{n_i}\cdot \mathbf{D}_{n_i,n_j}\cdot\vec{I}_{n_j},
    \label{eq:hamilton_cun}
\end{align}
where $\vec{S}$ denotes the electronic spin operator with $\beta_e$ the Bohr magneton, $\vec{I}_n$  the copper and nitrogen nuclear spins with $\gamma_n$  their respective gyromagnetic ratios.  The interaction strength $\mathbf{D}_{n_i,n_j}$  between nuclear spin $n_i$ and $n_j$ is calculated from magnetic dipolar interaction and is on the order of 100~Hz.

Since the electron spin Zeeman energy is much larger than the hyperfine interaction, the electron spin is well quantized by the external field, and hyperfine-induced electron spin flips are suppressed. We note that the quantization axis is determined by the effective Zeeman field, $\vec{B}_\text{eff} = B_0\mathbf{g}\cdot\vec{z'}\equiv B_\text{eff}\vec z$, and is therefore slightly misaligned with the external magnetic field direction due to $g$-factor anisotropy. The electron-nuclear Hamiltonian thus becomes $S_z\sum_n\vec{A}_{z,n}\cdot\vec{I}_n$. 
In our electron paramagnetic resonance (EPR) experiments, a microwave field, 
\begin{equation}
   {\cal H}_{\mu w} = \Omega_{\mu w} \cos(\omega_{\mu w} t + \phi_{\mu w}) S_{\perp},
    \label{eq:hamiltonian_pulse}
\end{equation}
probes the system transition frequencies. Here  $S_{\perp}$ is the electron spin operator  perpendicular to the quantization axis $z$, and $\Omega_{\mu w}$, $\omega_{\mu w}$, and  $\phi_{\mu w}$ are the amplitude, frequency, and phase of the microwave.
The echo-detected field sweep (EDFS)  signals for  CuPc:XPc powder samples (Fig.~\ref{fig:1}(c,d)) display four spin packets from Cu hyperfine couplings and nine additional sub-peaks from the $^{14}$N couplings.
This indicates that, 
due to the large hyperfine interaction, the microwave (amplitude  $\Omega_{\mu \mathrm{w}} = 15~\mathrm{MHz}$ ) only excites the nuclear spin manifolds which are resonant with the microwave.

\begin{figure*}[htbp]
\includegraphics[width=1\textwidth]{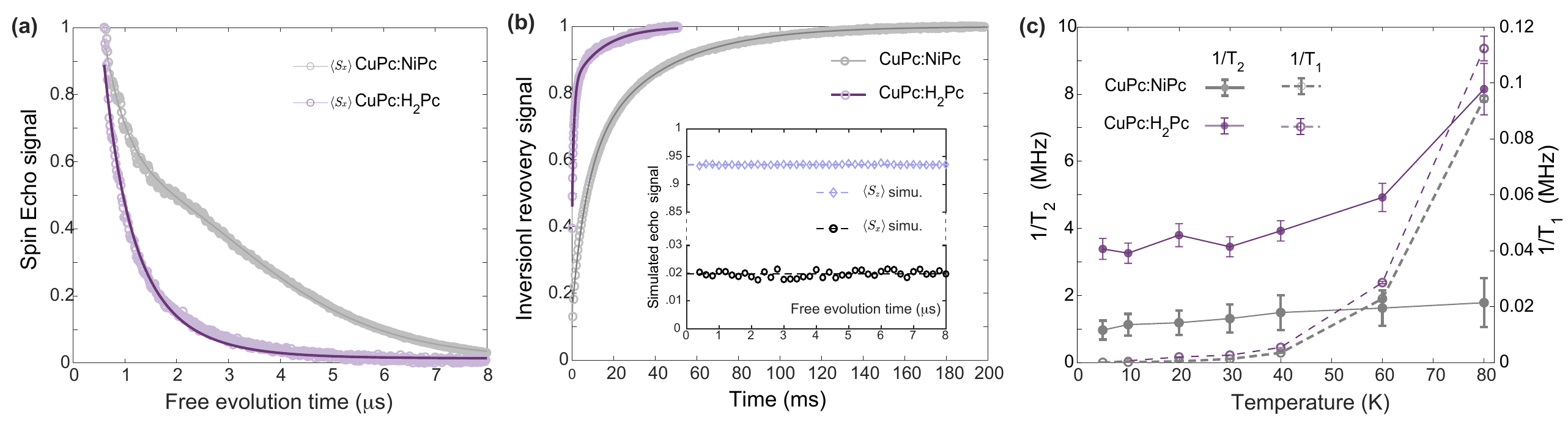}
 \caption{
 \textbf{CuPc relaxation time characterization.} 
(a) Normalized experimental spin-echo data for CuPc:NiPc and CuPc:H$_2$Pc at 5~K. The experimental decay times ($T_2$) are fitted to 1.0~$\mu$s and 0.3~$\mu$s for CuPc:NiPc and CuPc:H$_2$Pc, respectively.  
Inset of (b): simulated spin-echo decay of a single CuPc molecule (Eq.~\eqref{eq:hamilton_cun}), averaged over all molecular orientations with respect to the magnetic field (powder average). The longitudinal spin component $\langle S_z \rangle$ is significantly larger than the transverse component $\langle S_x \rangle$, indicating that only a limited fraction of the CuPc electron spin spectrum is driven into a superposition state by the microwave field. 
(b) Experimental inversion recovery measurements at 5~K. The experimental decay times ($T_1$) are fitted to 35~ms and 14~ms for  for CuPc:NiPc = 1:140 and CuPc:H$_2$Pc = 1:45, respectively. The data are shown only for $t > 20~\mu$s, corresponding to the spin-lattice relaxation component. (c) Temperature dependent of the transverse (solid line, $1/T_2$) and longitudinal (dashed line, $1/T_1$) decay rate.
}
\label{fig:2}
\end{figure*}

This approximation is supported by the exact numerical simulation of  a single CuPc molecule spin-echo evolution,  ensemble-averaged over different orientation of CuPc molecule. Under  a microwave field resonant with the transition at 3545~G, we simulate the system's time evolution assuming  the nuclear spins are in a fully mixed state, see Fig.~\ref{fig:2}(b). The results show that only a small fraction  $\langle S_{x} \rangle \ll \langle S_{z} \rangle$ of the electron spin population is excited by the drive.

The experimental transverse signal $\langle S_{x} \rangle$ (the EPR observable) decays much faster than the single-molecule simulation, even after powder averaging. 
This indicates that the strongly coupled nuclear spins in the CuPc molecule have only a minor effect on the electron spin coherence. 
Indeed, these nuclear spins  lie within the spin-diffusion barrier, so their flip-flop dynamics with other nuclear spins is effectively suppressed, and their hyperfine field can be treated as a static (frozen-core) contribution.~\cite{witzel2006quantum,guichard2015decoherence,cutsail2015advanced} 
As a result, the hyperfine coupling can be treated as a classical field that produces a fixed hyperfine shift, and we can evaluate the electron spin dynamics as occurring in distinct nuclear spin manifolds, labeled by $\iota$, each with transition frequency $\Delta_\iota$.


Having excluded the strongly coupled nuclear spins as the source of decoherence, we turn to analyzing other environmental sources that can contribute to transverse decoherence of the CuPc electron spin:
\begin{enumerate}
    \item Spin-lattice relaxation mediated by phonon, $R_{s-l}$
    \item Hyperfine coupling to a weakly interacting nuclear spin bath (e.g., protons), $R_n$
    \item Magnetic dipolar interactions with other CuPc electron spins, $R_e$.
\end{enumerate}
The total transverse decoherence rate will be the sum of these contributions 
listed in  Table~\ref{table:summary}. In the following sections, we analyze each contribution to evaluate their significance and provide insights into design principles for molecular spin qubits.


\begin{table*}[htbp]
\caption{Different transverse decoherence mechanisms of the electron spin of CuPc in XPc lattice.}
\label{table:summary}
\begin{center}
\renewcommand\arraystretch{1.3}
\begin{tabular}{cccc}
 \hline \hline
   \multirow{1}{2cm}{ \centering  ${1}/{T_{2}}$}&  \multirow{1}{6cm}{ \centering Source}&  \multirow{1}{2cm}{ \centering Effect} &  \multirow{1}{6cm}{ \centering Evidence}\\
 \hline
  \multirow{2}{2cm}{ \centering ${R_{Cu/N}}$}& \multirow{2}{6cm}{ \centering Nuclear spins coupled via Fermi contact interaction} & \multirow{2}{2cm}{ \centering Minor}& \multirow{2}{6cm}{ \centering Weak driving condition; numerical simulation (Section.~\ref{section:mechanics})}\\
  &&&\\
  \hline
   \multirow{2}{2cm}{ \centering ${R_{s-l}}$}& \multirow{2}{6cm}{ \centering Spin lattice interaction}&  \multirow{2}{2cm}{ \centering Minor}&
  \multirow{2}{6cm}{ \centering $T_2$ independent of temperature 
  $T_1>>T_2$ (Section.~\ref{section:spin_lattice})}\\
  &&&\\
  \hline
    \multirow{1}{2cm}{ \centering ${R_{n,cl}}$}&  \multirow{1}{6cm}{ \centering Classical nuclear spin bath}& \multirow{2}{2cm}{ \centering Minor} & \multirow{2}{6cm}{ \centering NOVEL experiment and CCE simulation (Section.~\ref{section:nuclear})}\\
  \cline{1-2}
   \multirow{1}{2cm}{ \centering ${R_{n,qu}}$}&  \multirow{1}{6cm}{ \centering Quantum nuclear spin bath} & & \\
  \hline
    \multirow{2}{2cm}{ \centering ${R_{e,FF}}$}& \multirow{2}{6cm}{ \centering Flip-flop process}& \multirow{2}{2cm}{ \centering Minor}& \multirow{2}{6cm}{ \centering Effectively reduced spin density due to hyperfine splitting (Section.~\ref{section:electron_ff})}\\
  &&&\\
  \hline
  \multirow{1}{2cm}{ \centering ${R_{e,ID}}$}& \multirow{1}{6cm}{ \centering Instantenous diffusion effect}& \multirow{2}{2cm}{ \centering Dominate} & \multirow{2}{6cm}{ \centering Spin bath modeling and CCE  simulation (Section.~\ref{section:electron_zz})}\\
  \cline{1-2}
  \multirow{1}{2cm}{ \centering $R_{e,SD}$}& \multirow{1}{6cm}{ \centering Spectral diffusion effect}&  & \\
  \hline
  \hline

\end{tabular}
\end{center}
\end{table*}

  
  

\section{Nuclear Spin Environment\label{section:nuclear}}

As discussed, the $^{63/65}\mathrm{Cu}$ and four $^{14}\mathrm{N}$ nuclear spins adjacent to the Cu(II) center mainly induce frequency shifts and negligible ESEEM due to hyperfine interactions, but do not substantially contribute to the spin echo decay.
More far away nuclear spins, instead, are not quenched and induce decoherence. Their coupling strength to the electron spin sets the characteristics of two distinct  nuclear  spin baths~\cite{wang2024digital,hernandez2018noise, reinhard2012tuning,taminiau2012detection}. The closest nuclear spins  to the Cu(II) center form a \textit{quantum} spin bath, which must be described by the coherent evolution of the hybrid electron-nuclear spin system~\cite{bradley2019ten}. In contrast, weakly coupled nuclear spins, whose dynamics are largely independent of the electron spin state, constitute a \textit{classical} spin bath~\cite{pham2016nmr}, which can be modeled by a stochastic magnetic field, whose fluctuations are captured by their power spectral density. Further details of this model are provided in the Supplementary Materials.

\begin{figure*}[htbp]
\includegraphics[width=1\textwidth]{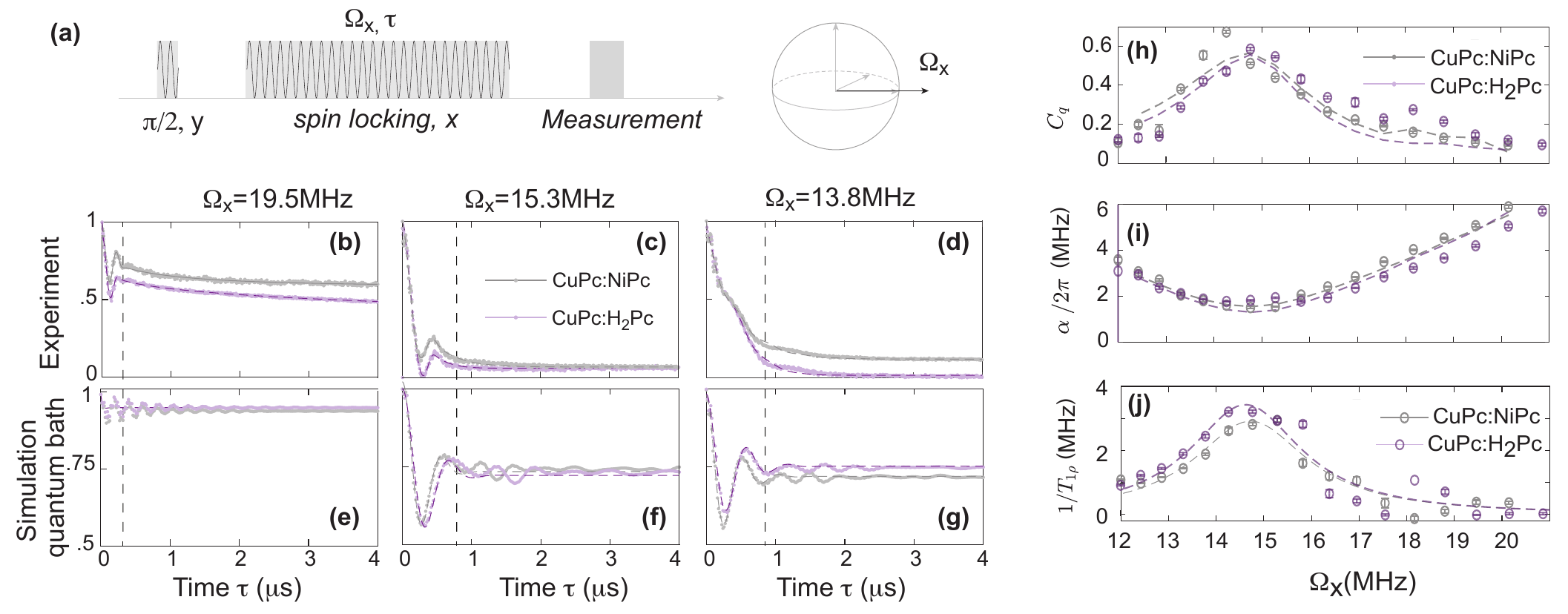}
\caption{\textbf{Spin locking response of CuPc to quantum and classical hydrogen spin baths.} 
    (a) Spin locking (NOVEL) pulse sequence. The locking pulse acts  as an effective static field along the x-axis in the rotating frame. 
    (b--c)Spatial distribution of $^1$H nuclear spins (purple dots) surrounding the CuPc electron spin (red dot with gray spin arrow), shown for 
     (b--d) Experimental spin locking signal and fit for CuPc:H$_2$Pc and CuPc:NiPc at varying $\Omega_x$. 
    (e--g) Simulated spin locking signal under a quantum bath only of hydrogen spins, at the same $\Omega_x$ as in (b--d). 
    (h-j) Signal oscillation amplitude $C_q$ and frequency $\alpha'$  (Eq.~\ref{eq:spinlockingfit}) as a function of $\Omega_x$ for CuPc:H$_2$Pc and CuPc:NiPc extracted from fits to the experimental data (circles) and from simulations (dashed lines.) 
    (l) Spin-locking classical decay rate $1/T_{1\rho}$ (Eq.~\ref{eq:P_classical_1}) CuPc:H$_2$Pc  and CuPc:NiPc(m) as a function of $\Omega_x$: Circles, wxperiment data; lines,   Lorentzian fitting.}
\label{fig:3}
\end{figure*}

The $^{1}\mathrm{H}$ nuclei exhibit both stronger hyperfine couplings and higher abundance than other nuclear spin species in the lattice, and therefore constitute the dominant contributors to the quantum-classical spin bath. 
In the H$_2$Pc matrix, the two additional protons located at the molecular center increase the overall $^{1}\mathrm{H}$ spin density by approximately 10\% compared to the NiPc matrix. Furthermore, the CuPc electron spin is  adjacent to the four center protons from two neighboring H$_2$Pc molecules (Fig.~\ref{fig:3}(b, c).) Comparing H$_2$Pc and NiPc can thus provide insights into the H spin role in decoherence.

We focus on a single on-resonance N-Cu manifold with frequency $\Delta_\iota$ and consider the reduced electron spin 
 Hamiltonian in the rotating frame  coupled to 
the hydrogen nuclear spin bath: 
\begin{align}
{\cal H}_n &={\cal H}_0+ {\cal H}_{e\text{-}n,q} + {\cal H}_{e\text{-}n,c}, \label{eq:hamilton_nuclearspin}
\end{align}
where the electron Hamiltonian ${\cal H}_0 = \delta_{\iota} S_z$ is set by $\delta_{\iota} = \Delta_{\iota_R} - \omega_{\mu w}$,  the detuning between the transition frequency $\Delta_{\iota_R}$ and the applied microwave frequency $\omega_{\mu w}$. 
The classical nuclear spin bath
\begin{align}
{\cal H}_{e\text{-}n,c} &= \gamma_e \tilde{B}_z(t) S_z\label{eq:hamiltonian_nc},
\end{align}
with $\gamma_e$  the gyromagnetic ratio of the electron, is described by   a time-dependent stochastic magnetic field, $\tilde{B}_z(t)$ generated by unpolarized hydrogen spins. The quantum component of the nuclear spin bath is given by
\begin{align}
{\cal H}_{e\text{-}n,q} &= \gamma_H{B}_0\sum_{n}I^n_z+ S_z \sum_{n}\sum_{\sigma = x,y,z} A_{z\sigma}^{n} I_{\sigma}^{n} , \label{eq:hamiltonian_nq}
\end{align}
where $\gamma_H$  is the hydrogen gyromagnetic ratio and $A_{z\sigma}^n$ is the hyperfine coupling tensor between the electron spin and the $n$-th $^{1}\mathrm{H}$ nuclear spin. At the experimental field $B_0 = 3545$~G the Larmor frequency is $\omega_H\approx15.09$~MHz.

To selectively probe the hydrogen nuclear spin bath, we performe a spin locking experiment~\cite{henstra1988nuclear,henstra1988nuclear}, whose sequence is shown in Fig.~\ref{fig:3} (a). The electron spin is first rotated to the x axis by a  $\pi/2$ $S_y$ pulse and then we apply a
continuous microwave drive at the same frequency with a 90$^\circ$ phase shift  to lock the electron spin. In the rotating frame (within the rotation wave approximation), the drive  Hamiltonian,
\[ {\cal H}_{\text{locking}} = \Omega_x S_x,\]
 is added to the system Hamiltonian in Eq.~\eqref{eq:hamilton_nuclearspin}.  This spin-locking field induces a dressed-state energy splitting of the electron spin $\Omega_r=\sqrt{\Omega_x^2+\delta_\iota^2}$, which protects against decoherence while enabling resonant polarization transfer to nuclear spins when this splitting matches the nuclear Larmor frequency. This protocol --also known as  Nuclear Orientation Via Electron Spin Locking (NOVEL)-- thus enables probing the environmental proton spins as done in quantum noise spectroscopy~\cite{suter2016colloquium,cywinski2014dynamical,yan2013rotating}.  

The spin coherence as a function of the the spin locking pulse duration is shown in Fig.~\ref{fig:3}(d--f) for exemplary driving strengths. When varying the driving strength $\Omega_x$, the electron spin  decoherence rate  reaches a maximum at the resonant condition $\Omega_r\sim\omega_H=15.09$~MHz. (See additional experimental results in the Supplementary Material).
We observe two distinct components: a fast, damped oscillation within the first 2~$\mu$s and a slower exponential decay tail. 
We ascribe each feature to the  quantum and classical spin baths, respectively, as we can prove with a simple model and numerical simulations (Fig.~\ref{fig:3}(g--i)) 


The oscillation arises from coherent polarization exchange between the electron spin and the surrounding quantum spin bath. The presence  of multiple polarization transfer pathways between the electron spin and nuclear spins with varying coupling strength and ensemble averaging over all possible detuning $\delta_\iota$ and molecular orientations leads to a damping of the oscillations.  
We can reproduce the key features of the experimental oscillation, such as the frequency and damping rate, by simulating  the electron spin dynamics under the interaction with a pure quantum spin bath. The  electron spin polarization under coupling to a quantum spin bath is well described by :
\begin{equation}
    P_q(t) = (1 - C_q) - C_q e^{-\Gamma_q t} \cos(\alpha t),
    \label{eq:spinlockingfit}
\end{equation}
where $C_q$ denotes the polarization transfer amplitude, $\Gamma_q$ the damping rate , and $\alpha$  the oscillation frequency. A detailed analytical and numerical analysis of the spin-locking evolution is shown  in the Supplementary Material.

The slower decay tail, which is captured by a simple exponential  
\begin{equation}
    P_c(t) = e^{-t / T_{1\rho}},
    \label{eq:P_classical_1}
\end{equation}
arises from the classical spin bath~\cite{henstra2008theory,yan2013rotating,wang2020coherence} which sets the spin-locking relaxation time, $T_{1\rho} \propto \frac{2}{\gamma_e^2 S(\Omega_r)}$, to be proportional to the spectral density $S(\Omega_r)$ of the fluctuating magnetic field at the dressed-state splitting $\Omega_r$.

The total EPR signal after spin locking  thus includes both the quantum and classical contributions, 
\(    P_{\mathrm{spin-locking}}(t) = P_q(t) \, P_c(t)\).
Fitting the data to this expression, we extract the relevant parameters 
as shown in Fig.~\ref{fig:3}(j--l).

The fitted quantum spin bath parameters ($C_q$, $\alpha$) for H$_2$Pc and NiPc are nearly identical, consistent with the simulation results. The peak in $C_q$ and the dip in $\alpha$, corresponding to the maximum polarization transfer amplitude and the transfer rate, respectively, occur at the resonance condition  $\Omega_r\approx \omega_H$.

The spectral densities of the classical  hydrogen spin bath, given by $\gamma_e S(\omega) \propto  2 / T_{1\rho}$, are also similar in the two samples. H$_2$Pc exhibits a slightly higher peak amplitude (approximately 10\%) near the Larmor frequency, consistent with its higher hydrogen density.
The  linewidth of $S(\Omega_r)$ is on the order of 2~MHz, significantly broader than the intrinsic linewidth expected for a hydrogen nuclear spin bath, typically in the kilohertz range. This indicates that the $T_{1\rho}$'s linewidth is dominated by the inhomogeneous broadening of the EPR transitions, which  
can reach  8~MHz due to the $g$-factor anisotropy averaging in the powdered CuPc sample.
Due to the EPR transition broadening, the Hartmann-Hahn condition $\Omega_r = \omega_{\textrm{H}}$ is satisfied over a larger frequency range (here  
approximately 2~MHz) as already observed 
in similar molecular systems~\cite{mathies2016pulsed,jain2017off}.

The results of the spin-locking experiments suggest that the additional hydrogen nuclear spins in the H$_2$Pc molecule have a minimal effect on  the combined quantum and classical spin bath. This observation also implies that other possible substitutions of the X molecule are also likely to have a limited impact on the spin bath composition. 

The spin-locking decay show that the role of the H in decoherence only becomes large in a relatively narrow range close to the resonance condition. We thus also expect that the H spin will not dominate the $T_2$ echo decay time (unless pulse spacings were close to the resonant condition, which is not the case in our experiments.)
Still, we cannot experimentally isolate the contribution of nuclear spins --and in particular hydrogen nuclear spins-- in the 
spin echo ($T_2$) decoherence. We thus resort to numerical simulations using  the cluster correlation expansion (CCE) method~\cite{yang2008quantum,onizhuk2021pycce}.
We use the results obtained from spin locking measurements and modeling to incorporate  both quantum ($T_{n,\mathrm{qu}}$) and classical ($T_{n,\mathrm{cl}}$) components of the nuclear spin environment in the CCE model. 
By directly simulating the many-body dynamics (truncated to computationally accessible cluster sizes), CCE calculates   the total decoherence rate:
\[
\frac{1}{T_{n,\mathrm{cce}}} = \frac{1}{T_{n,\mathrm{qu}}} + \frac{1}{T_{n,\mathrm{cl}}}.
\]
The simulated decoherence rates due to nuclear spins for H$_2$Pc and NiPc are $T_{n,\mathrm{cce}}^{-1} = 90$~kHz and 80~kHz, respectively. These rates are significantly smaller than the experimentally observed echo decay rates $T_2^{-1}$, which lie in the MHz range. This discrepancy indicates that decoherence mechanisms beyond the nuclear spin bath contribute substantially to spin echo decay. Indeed, we do not expect a static (inhomogeneous) broadening of the linewidth to contribute to spin echo decay, since the $\pi$-pulse effectively refocuses static detuning effects. Spin echo $T_2$ measurements are instead  sensitive to dynamic fluctuations on the timescale of the echo duration.

Combining insights from both spin-locking experiments and CCE simulations, we conclude that the presence of additional hydrogen nuclear spins at the center of the XPc molecule does not significantly alter the nuclear spin environment. Furthermore, the nuclear spin bath contributes only a minor portion to the experimentally observed $T_2$ decoherence.

\begin{figure}[htbp]
\includegraphics[width=0.46\textwidth]{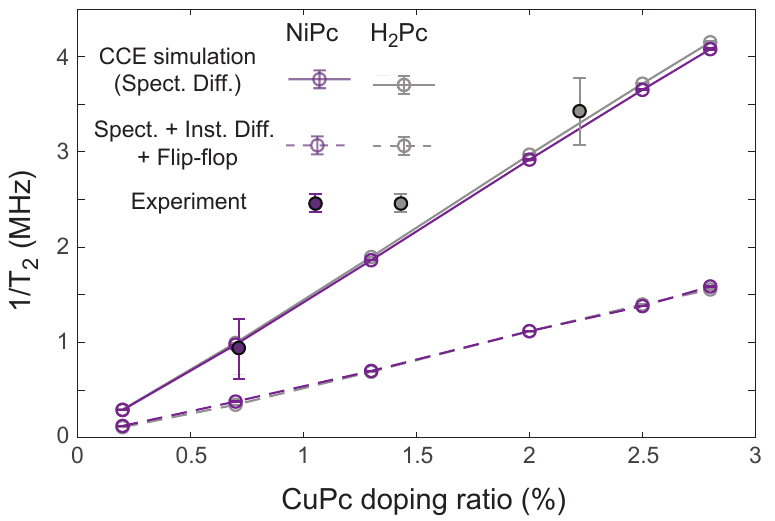}
\caption{\textbf{CCE results and electron--electron spin interaction effects.} The $T_2$ decay rates obtained from CCE simulations are shown as gray (CuPc:NiPc) and purple (CuPc:H$_2$Pc) curves, at varying CuPc doping percentages. The black dashed line represents the simulated total decoherence rate from electron spin bath interactions, incorporating \textit{flip-flop interactions}, \textit{instantaneous diffusion}, and \textit{spectral diffusion} discussed in Section.~\ref{section:electron}. Experimental data measured at doping ratios of CuPc:NiPc = 1:140 and CuPc:H$_2$Pc = 1:45 are also plotted, showing good agreement with the simulated overall electron spin bath effect.
\label{fig:4}}
\end{figure}

\section{Electron Spin Environment\label{section:electron}}
Although much less dense, electronic spins
can greatly contribute to decoherence, due to their stronger interactions. 
In an CuPc:XPc crystal, the electron spin bath predominately consists of the electron spins from all the other CuPc spins~\cite{warner2013potential,sushkov2014magnetic,escalera2019decoherence}, giving rise to a complex many-body system.

Despite the apparent lack of distinction between ``system'' and ``environment'', we can still consider how a probe CuPc \textit{central} spin  is  
affected by a spin bath consisting of the other CuPc molecules. According to the simulation of the the dynamics in a single molecule in Section.~\ref{section:mechanics}, where the strongly coupled nuclear spins ($Cu$, $N$) are effectively frozen and only contribute to frequency shift.
In particular, we can treat the nuclear spins as classical (and frozen) and 
we can focus on a central effective spin-\(1/2\) (representing the spin packet that is resonantly driven in the spin echo sequence, labeled by \(\iota_R\)) coupled to a bath of electronic spins each with their effective frequency $\Delta_\iota$,

The Hamiltonian for \(N\) electronic spins (\(N\) CuPc molecules within a finite volume) can be written as
\begin{equation}
\begin{aligned}
    {\cal H}_e' = &\ \Delta_{\iota_R} S_z^{\iota_R} + \sum_{k=1}^N \vec{S}^{\iota_R} \cdot \mathbf{D}^k \cdot \vec{S}^k \\
    &+ \sum_{k=1}^N \left( \Delta_{\iota_k} S_z^k + \sum_{k'=1}^N \vec{S}^{k'} \cdot \mathbf{D}^{k'k} \cdot \vec{S}^k \right),
\end{aligned}
\label{eq:hamilton_electronspin_2}
\end{equation}
where \(k\) indexes individual electron spin sites each with frequency $\Delta_{\iota_k}$ set by the (random) nuclear spin state. 
We note that here are $2M$ nuclear spin packets because the $\beta$-phase CuPc-XPc crystal contains two distinct orientations of CuPc molecules in the lattice (illustrated as up-tilted and down-tilted columns in Figure~\ref{fig:1}), and each orientation exhibits a distinct set of \(M=36\) hyperfine-shifted spin-\(1/2\) transitions. Evolution under this Hamiltonian is then averaged over the CuPc  positions and the intra-molecule nuclear spin  states at thermal equilibrium.

 We note that due to the large Zeeman energy in the chosen magnetic field, the dipolar interaction $\vec{S}^{\iota_R}\cdot \mathbf{D}^k\cdot \vec{S}^k$ and $\vec{S}^{k'} \cdot \mathbf{D}^{k'k} \cdot \vec{S}^k$ in Eq.~\eqref{eq:hamilton_electronspin_2} reduces to its secular part with two distinct components, a transverse term \(D_{\perp}(S_xS_x + S_yS_y)\) and a longitudinal term $D_{zz}S_zS_z$. In the following, we  quantitatively investigate the coherence properties of the central spin under these components.

\subsection{Flip-Flop Processes\label{section:electron_ff}}
The transverse component of the magnetic dipolar interaction induces flip-flop processes. By using the standard ladder operators \(S_{\pm}\)  to rewrite  it,  (\(D_{\perp}(S_xS_x + S_yS_y)\equiv D_{\perp}(S_+S_- + S_-S_+)\)), makes it clear that this interaction drives transitions \(\ket{\uparrow\downarrow} \rightarrow \ket{\downarrow\uparrow}\) between spin pairs, leading to polarization transfer that causes decoherence.

The flip-flop rate between two  spins with  energy mismatch $\Delta$ and transverse coupling strength $D_{\perp}$ is
\begin{equation}
\Gamma_{\text{ff}}(\Delta) = \frac{D_{\perp}^2}{\sqrt{\Delta^2 + D_{\perp}^2}}.
\end{equation}
To evaluate the flip-flop effect we consider again a central probe spin interacting with a large bath of electronic spins in a thermal state. 
The spin bath contains \(M\) equally populated  nuclear spin manifolds, setting the frequency of each electronic spin,  \(\Delta_{\iota}\). The total flip-flop rate experienced by the central spin is the sum of contributions from all of the M groups, 
\begin{equation}
\Gamma_{\mathrm{flip\text{-}flop}} 
= \sum_{\iota}  \frac{\langle D_{\perp,\iota}^2\rangle}{\sqrt{(\Delta_{\iota_R} - \Delta_{\iota})^2 + \langle D_{\perp,\iota}^2\rangle}},
\end{equation}
where $\langle D_{\perp,\iota}^2\rangle$ is second moment of the dipolar couplings of a bath with spin density \(n_e^\iota=n_e/M\): $\sqrt{\langle D_{\perp,\iota}^2\rangle} = \frac{\mu_0\hbar}{4\pi} \gamma_e^2  \sqrt{\frac{8}{105}} \frac{n_e}M$.
 The contribution of flip-flops  to the transverse relaxation rate can be approximated as 
\(R_{\mathrm{flip\text{-}flop}} = \frac{1}{2} \Gamma_{\mathrm{flip\text{-}flop}}.\)
For CuPc:H\textsubscript{2}Pc at a 1:49 doping ratio and CuPc:NiPc at 1:140 we estimate $ \Gamma_{\mathrm{flip\text{-}flop}} \approx 0.11~\mathrm{MHz} $ and \(\approx 0.03~\mathrm{MHz}\), respectively, indicating that only spins from the same manifold as $\iota_R$ contribute. These values are significantly lower than the measured spin echo dephasing rates, which lie in the MHz range. 
These observations allow us to simplify our subsequent model for the electronic spin bath and focus on the effects of the longitudinal dipolar term. In doing so, we can assume that the central spin will not undergo a spin flip in the time it takes it to decohere due to the dipolar longitudinal term contribution. We remark that this small effect of the flip-flop can be attributed to the effective reduced density, $n_e/M$, of the spin bath that arises from the resolved nuclear spin manifolds.

\subsection{Dephasing from $S_z$ Interaction\label{section:electron_zz}}

The longitudinal term in the dipolar interaction,   \(\sim~\!D_{\parallel} S_z\), contributes to dephasing and thus plays a crucial role in transverse decoherence. 
Its effects in spin echo experiments depends on whether the bath spins are themselves driven by 
the microwave pulses. Then we can describe the spin probe to be interacting with two distinct sets of CuPc spins: ``\textit{A} spins'', which  are driven by the control microwave field because their transition frequencies are near  resonance; and ``\textit{B} spins'', which are unaffected by the microwave excitation due to their energy detuning  due mostly to hyperfine interactions.   
$A$~spins give rise to instantaneous diffusion while $B$ spins to spectral diffusion, two distinct dephasing mechanism that   we analyze next.

\subsubsection{Instantaneous Diffusion Effect}

Instantaneous diffusion (ID)  arises from longitudinal dipolar interactions between spins that are simultaneously flipped by a microwave pulse, that is,  $A$ spins. 
In a spin echo, the longitudinal coupling \(D_\parallel S_z S_z\) between two $A$ spins remains invariant under the  (resonant) \(\pi\)-pulse. 
As a result,
the phase accumulated due to $D_\parallel^k$ is not refocused and   results in incoherent dephasing among the $A$ spins.

While this is a large effect, not mitigated by the echo, it is somewhat alleviated by the effective smaller density of the $A$ spins, $n_e/M$. 
 Then, the instantaneous diffusion dephasing rate in  our CuPc:XPc samples can be quantitatively estimated to be
~\cite{schweiger2001principles}:
\begin{equation}
    R_{\mathrm{ID}} = \frac{n_e}{M}\cdot \frac{4\pi^2}{9\sqrt{3}} \cdot \frac{\mu_0 \gamma_e^2 \hbar}{4\pi}.
\end{equation}
This estimate yields \(R_{\mathrm{ID}} \approx 0.6\,\text{MHz}\) for CuPc:NiPc at a mixing ratio of 1:140, and \(R_{\mathrm{ID}} \approx 1.88\,\text{MHz}\) for CuPc:H\textsubscript{2}Pc at 1:49


\subsubsection{Spectral Diffusion Effect}
The $A$-$B$ longitudinal dipolar interaction, $\sim~D_\parallel S_zS_z$, is ideally canceled by the spin echo $\pi$ pulse on the A spin. However, the B spin dynamics leads to imperfect cancellation. This dynamics changes the effective local field felt by the A spin, thus leading its frequency to ``diffuse'' during the evolution, which causes decoherence
~\cite{schweiger2001principles}. 

In contrast to the electron-nuclear spin interaction model, the dipolar interactions among bath spins ($B$ spins), see second line of Eq.~\ref{eq:hamilton_electronspin_2},  are of the same order of magnitude as the interactions between the central spin ($A$ spin) and the bath spins.
This fast dynamics 
contribute to the decoherence of the central spin during a spin echo sequence. 

To quantitatively evaluate the many-body dynamics governed by this Hamiltonian, we employ the cluster-correlation expansion (CCE) method for an electronic spin bath. For each simulation instance, bath spins are randomly placed at molecular sites of XPc within the $\beta$-phase CuPc:XPc crystal lattice in a finite simulation volume. Each site is assigned a transition frequency randomly drawn from the simulated hyperfine spectrum corresponding to its specific molecular orientation.

Spin echo dynamics is then simulated for each generated spin configuration. Ensemble averaging is performed by repeating the initialization of the bath over many realizations until convergence is reached. Since the experiment is conducted on a powder sample, the final simulated signal is obtained by averaging over all possible crystal orientations. Additional details regarding simulation parameters and convergence criteria are provided in the Supplementary Material.

The CCE simulation  yields a spectral diffusion-induced decoherence rate of \(R_{\mathrm{SD}} = 1~\mathrm{MHz}\) for CuPc:NiPc = 1:140 and \(R_{\mathrm{SD}} = 1.5~\mathrm{MHz}\) for CuPc:H\(_2\)Pc = 1:49. 

By combining these results with the previously estimated contribution from A-A spin interactions (instantaneous diffusion), we obtain a complete prediction of the transverse decoherence rate as a function of CuPc:XPc doping ratio, as shown in Fig.~\ref{fig:4}. The predicted values agree well with experimental measurements, confirming that at these densities decoherence is dominated by the electron spin bath effects. 

Furthermore, the approximately linear dependence of the transverse decoherence rate on the CuPc concentration suggests that \(T_2^{-1}\) can serve as a metric for electron spin density across a wide doping range.


\section{Spin-Lattice Interaction\label{section:spin_lattice}}
We finally consider spin-lattice interactions, mediated by spin-orbit coupling, which lead to energy relaxation of the electron spin system and are characterized by the longitudinal relaxation time $T_1$. These interactions also contribute to spin dephasing at a rate on the order of $R_{\mathrm{sl}} = 1/(2T_1)$. 


To quantify spin-lattice relaxation, we performed saturation-recovery measurements to extract $T_1$ over a broad temperature range for both CuPc:NiPc and CuPc:H$_2$Pc samples. Immediately following the saturation pulse train, spin diffusion via flip-flop processes between on-resonance and off-resonance CuPc molecules occurs during the initial stage of free evolution, typically on the order of $\sim$10~$\mu$s (see Section~\ref{section:electron}). Subsequently, a slower relaxation process dominated by spin-lattice interactions takes place.



As shown in Fig.~\ref{fig:2}, at low temperature (5\,K), the transverse decoherence time $T_2$ is significantly shorter than $T_1$, indicating that temperature-dependent spin--lattice relaxation contributes minimally to $T_2$ in this regime. 
Furthermore, the temperature dependence of $T_2$,  shows that $1/T_2$ varies only weakly with temperature below $\sim$40\,K. 
These results indicate that in the low-temperature regime, decoherence is dominated by temperature-independent mechanisms, such as the spin--spin interactions analyzed above.


\section{Summary and Outlook}




In this work, we have demonstrated the potential of molecular spin qubits by performing a comprehensive investigation of the decoherence mechanisms in copper(II) phthalocyanine (CuPc) diluted into diamagnetic phthalocyanine matrices.
Through systematic experimental studies and quantitative modeling grounded in quantum information science, we identified and characterized all major decoherence channels (summarized in Table.~\ref{table:summary}).
By explicitly distinguishing between strongly hyperfine-coupled nuclei (Cu, N) and weakly coupled nuclei (H), we established that neither makes a significant contribution to $T_2$, in contrast to common assumptions in earlier studies.
Our results show that longitudinal electron--electron dipolar interactions, manifested through instantaneous and spectral diffusion, constitute the primary limitation to coherence.
This finding was validated by direct comparison between cluster correlation expansion simulations and experimental spin-echo dynamics, which consistently revealed that dipolar couplings dominate even at moderate spin densities.
The agreement between simulation and experiment not only confirms the reliability of the framework, but also allows us to accurately estimate the electron spin density from the dipolar coupling-induced decoherence, a parameter that has  been previously difficult to quantify.
Hence we not only demonstrated dipolar-limited coherence but provided a systematic, predictive methodology for disentangling decoherence channels in molecular spin ensembles.
Our approach overcomes   oversimplified assumptions of earlier models and establishes a transferable framework for evaluating spin-bath contributions across different host environments.
Importantly, we find that variations in nuclear composition of the diamagnetic host (XPc) have little effect on $T_2$, underscoring that electron--electron interactions will remain the universal bottleneck in these systems.
The framework presented here can be applied broadly to other classes of molecular qubits and solid-state spin systems, guiding the rational design of high-coherence molecular platforms.
These insights provide design rules for tailoring spin density and interaction strengths, thereby advancing the development of scalable quantum devices, molecular spin-based sensors, and hybrid quantum materials.

\acknowledgments
We thank Prof. Tom Wenckebach for insightful comments. This work has been supported by Honda Research Institute USA Inc., and by the NIH through grants GM132997 and S10OD028706.

\bibliographystyle{apsrev4-1}
\bibliography{CuPc_EPR_ref} 

\end{document}


\begin{center}
{\bf \large{Supplemental Materials:}}\\
{\bf \Large{Exploring the mechanisms of transverse relaxation of copper(II)-phthalocyanine spin qubits}}
\end{center}
\author{Boning Li}\thanks{These authors contributed equally.}
    \affiliation{Department of Physics, Massachusetts Institute of Technology, Cambridge, MA 02139, USA}
    \affiliation{Research Laboratory of Electronics, Massachusetts Institute of Technology, Cambridge, MA 02139, USA}
    
\author{Yifan Quan}\thanks{These authors contributed equally.}
    \affiliation{Department of Chemistry and Francis Bitter Magnet Laboratory, Massachusetts Institute of Technology, Cambridge, MA 02139, USA}
\author{Xufan Li}\thanks{These authors contributed equally.}
   \affiliation{Honda Research Institute USA, Inc., San Jose, CA 95134, USA}
\author{Guoqing Wang}
    \affiliation{Research Laboratory of Electronics, Massachusetts Institute of Technology, Cambridge, MA 02139, USA}
    \affiliation{Department of Physics, Massachusetts Institute of Technology, Cambridge, MA 02139, USA}
\author{Robert G Griffin}
    \affiliation{Department of Chemistry and Francis Bitter Magnet Laboratory, Massachusetts Institute of Technology, Cambridge, MA 02139, USA}
    \author{Avetik R Harutyunyan}\email{aharutyunyan@honda-ri.com}
    \affiliation{Honda Research Institute USA, Inc., San Jose, CA 95134, USA}
    \affiliation{Department of Nuclear Science and Engineering, Massachusetts Institute of Technology, Cambridge, MA 02139, USA}
\author{Paola Cappellaro}\email[]{pcappell@mit.edu}
    \affiliation{Department of Physics, Massachusetts Institute of Technology, Cambridge, MA 02139, USA}
    \affiliation{Research Laboratory of Electronics, Massachusetts Institute of Technology, Cambridge, MA 02139, USA}
    \affiliation{Department of Nuclear Science and Engineering, Massachusetts Institute of Technology, Cambridge, MA 02139, USA}


\maketitle

\tableofcontents
\section{Material synthesis method\label{appendix:growcupc}}
The CuPc and Ni- or H$_2$-diluted CuPc crystals were synthesized using a vapor transport method. Precursors of raw CuPc powder (Sigma-Aldrich, $>$99\%) or a mixture of CuPc and NiPc/H$_2$Pc powders (Sigma-Aldrich, 98\%) at specific weight ratios were loaded into one end of a 0.5"-diameter, 6"-long glass tube, which was subsequently vacuum-sealed with a base pressure of approximately $10^{-3}$ Torr. The precursor powder in the glass tube was then heated in a tube furnace equipped with a 1" quartz tube at a temperature of approximately 530 $^\circ C$. The Pc crystals were grown at the other end of the glass tube over a growth time of 20 minutes and were collected from the tube wall using a wood stick after cooling down.

The precise concentration of Cu in the diluted CuPc crystals was determined by inductively coupled plasma-mass spectrometry (Thermo Scientific iCAP RQ ICP-MS) and 45Sc, 89Y, 159Tb and 209Bi as internal standards. Ni or H$_2$-diluted CuPc crystals were dissolved by adding 0.75 mL concentrated HNO$_3$ and 0.25 mL concentrated HCl to prepare a stock solution. The ICP-MS sample was prepared by diluting 0.10 mL of the stock solution with 9.9 mL deionized water. For the mixed precursor of CuPc + NiPc and CuPc + H$_2$Pc with a weight ratio of 1:100, the actual molar ratio of Cu:Ni and CuPc:H$_2$Pc was determined by ICP-MS measurement to be 1:45 and 1:140, respectively.
\section{CuPc:NiPc and CuPc:H$_2$Pc Phase Determination}
\begin{figure}[htbp]
    \centering
    \includegraphics[width=0.7\linewidth]{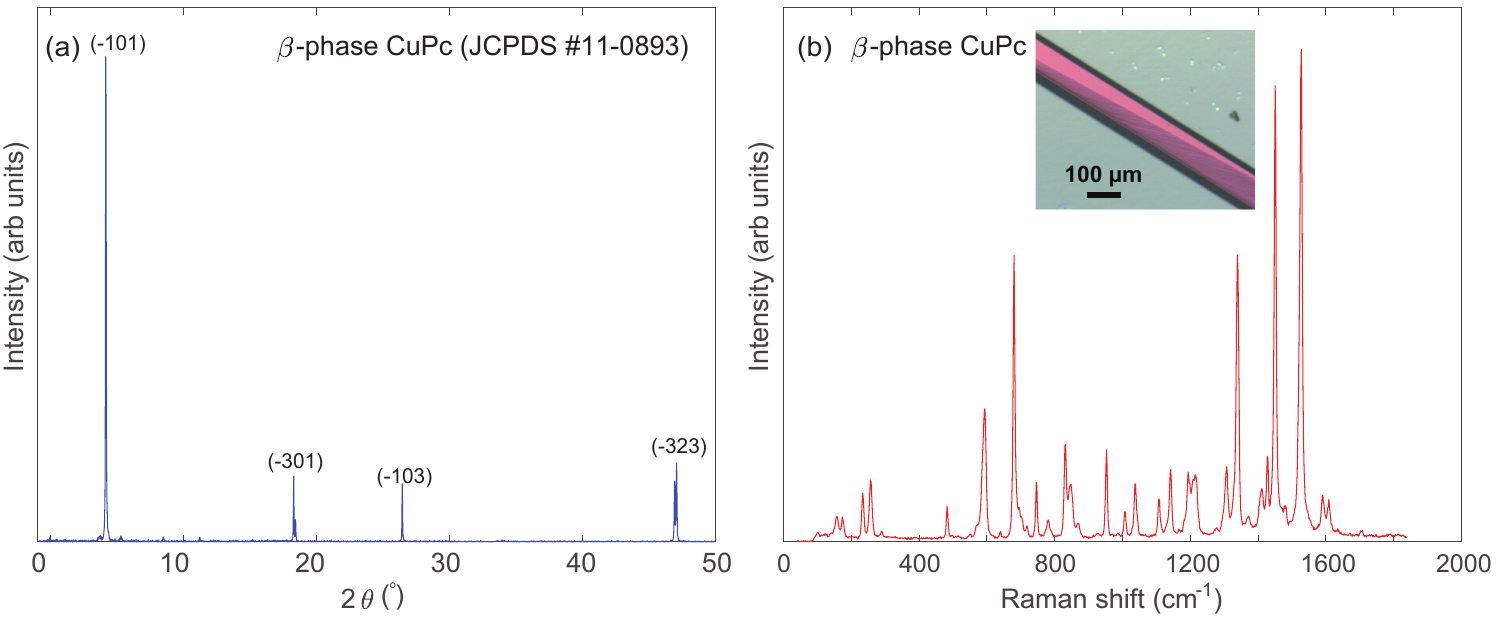}
    \caption{\textbf{$\beta$-CuPc:XPc characterization.} 
    (a) Powder XRD pattern of CuPc crystals. The diffraction peaks match well with JCPDS \#11-0893 in the ICDD database, corresponding to $\beta$-phase CuPc with a P2$_1$/c space group.  
    (b) Raman spectrum of the CuPc crystal, showing characteristic $\beta$-phase peaks~\cite{zou2018controllable}. Inset: optical microscope image of a single $\beta$-phase CuPc crystal.}
    \label{figS:raman_xrd}
\end{figure}

The CuPc:XPc (X = Ni or H$_2$) crystals were characterized using X-ray diffraction (XRD) and Raman spectroscopy. XRD measurements were performed on a \textit{Bruker AXS D8 Advance A25} system with Cu K$\alpha$ radiation ($\lambda$ = 1.5418~\AA). Raman spectra were collected with a \textit{Renishaw inVia Raman Microscope} equipped with an 1800~grooves/mm grating and a 532~nm excitation laser (1~mW power). The results, shown in Fig.~S\ref{figS:raman_xrd}, confirm that both samples exhibit the $\beta$-phase crystal structure.  

\section{Additional $T_1$ and $T_2$ Measurements}
\begin{figure}[htbp]
    \centering
    \includegraphics[width=0.7\linewidth]{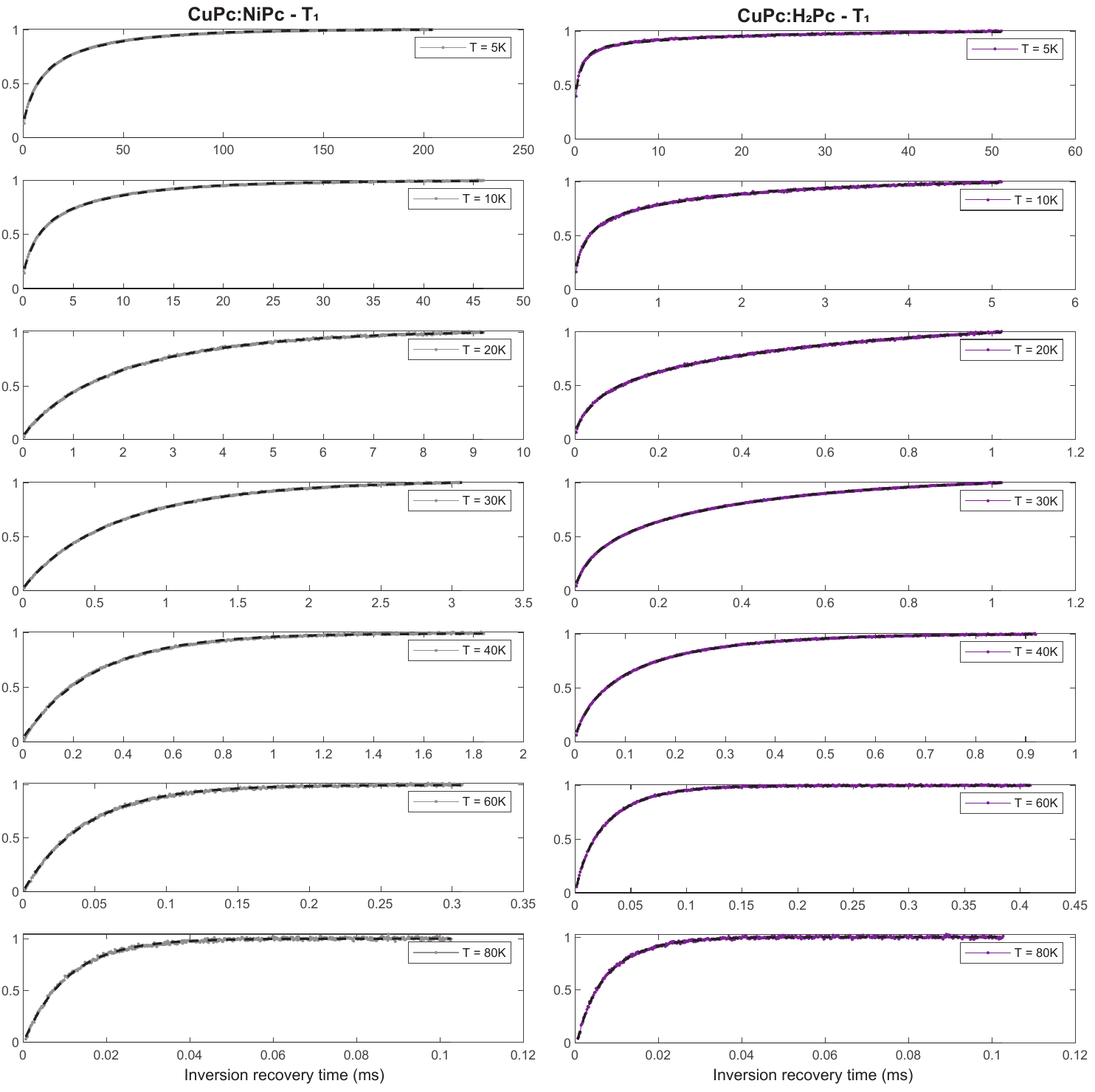}
    \caption{Inversion-recovery ($T_1$) measurements for CuPc:NiPc (right) and CuPc:H$_2$Pc (left) at $B_0=3545~\mathrm{G}$. Data at multiple temperatures are fitted to a bi-exponential recovery model,
$M_z(t)=1 - c_1 e^{-t/T_{\mathrm{s\text{-}l}}} - \big(1-c_1\big)e^{-t/T_{\mathrm{s\text{-}s}}}$,
with $0\le c_1\le 1$. As discussed in the main text, the fast component $T_{\mathrm{s\text{-}s}}$ is attributed to spin--spin processes (e.g., flip-flop with the off-resonant spin manifolds), whereas the slow component $T_{\mathrm{s\text{-}l}}$ reflects spin--lattice relaxation.}
    \label{figs:epr_result_T1}
\end{figure}

\begin{figure}[htbp]
    \centering
    \includegraphics[width=0.7\linewidth]{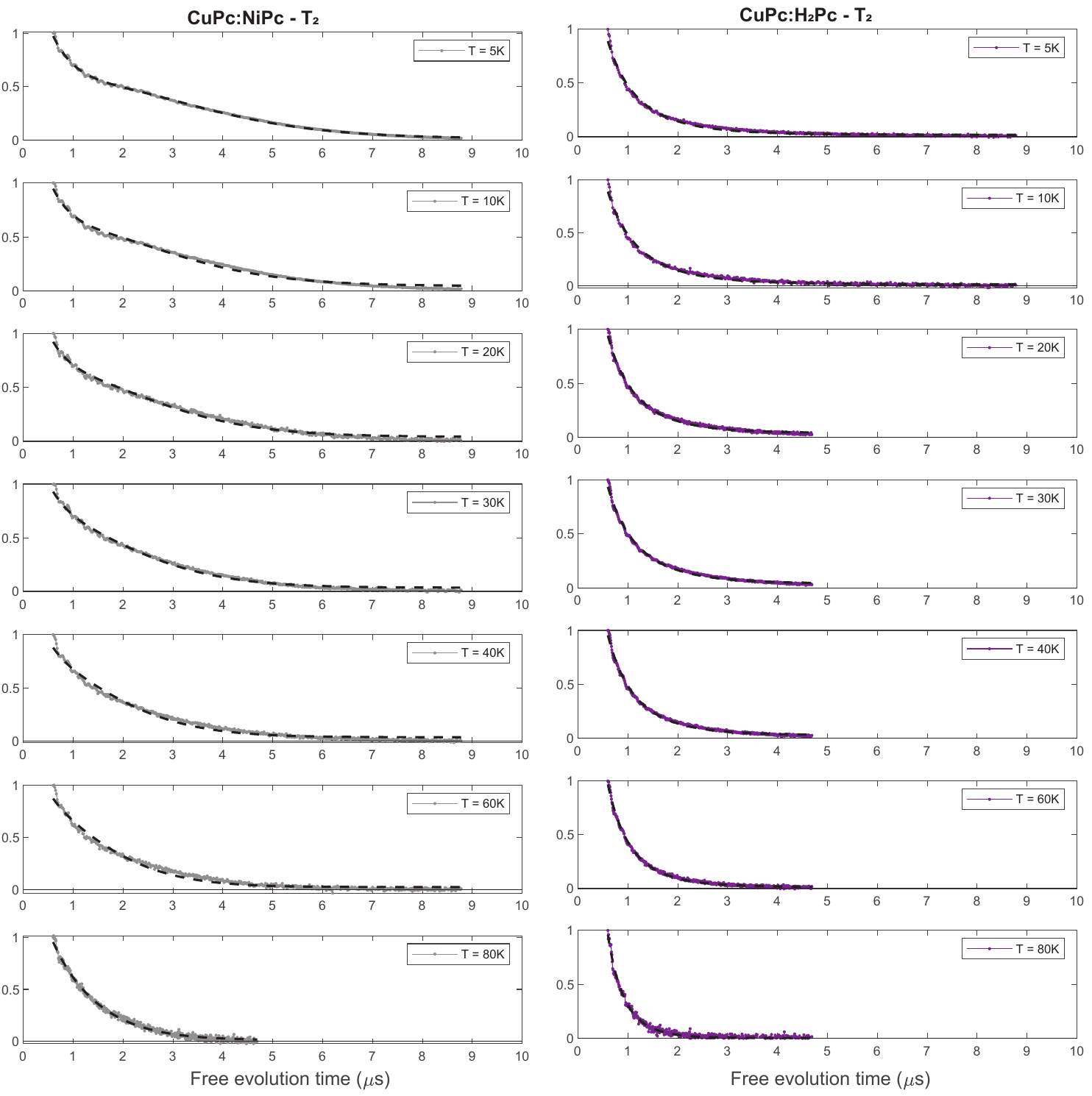}
    \caption{Spin–echo ($T_2$) measurements for CuPc:NiPc (right) and CuPc:H$_2$Pc (left) at $B_0=3545~\mathrm{G}$. Data at multiple temperatures are fitted with
$M_x(t)=\big[c_0+c_1\cos\!\big(2\pi f_b t+\phi\big)\big]\,e^{-t/T_2}$.
The phenomenological modulation, a form commonly used in previous molecular spin echo experiments~\cite{warner2013potential,atzori2016room}  does not bias the extracted $T_2$, which is set by the envelope $e^{-t/T_2}$. Numerical simulations reproduce the modulation and attribute it to hyperfine coupling between the CuPc electron spin and the four nearest $^{14}\mathrm{N}$ nuclei.}
    \label{figs:epr_result_T2}
\end{figure}

We measured the $T_1$ and $T_2$ times of CuPc in both NiPc and H$_2$Pc matrices over a temperature range of 5-100~K, as shown in Fig.~S\ref{figs:epr_result_T1} and S\ref{figs:epr_result_T2}. The corresponding longitudinal and transverse decay rates ($1/T_1$ and $1/T_2$) are plotted as a function of temperature in Fig.~2 (c) in the maintext..  
The results are consistent with previous studies, which report that longitudinal relaxation is strongly temperature-dependent, while the transverse decoherence rate remains nearly constant at low temperature. This indicates that transverse decoherence is dominated by temperature-independent mechanisms, such as spin-spin interactions, as analyzed in the main text.
\section{Simulation of the Spectral density of electron spin of CuPc \label{appendix:spectrum}}

We simulated the EPR spectrum of CuPc molecules under an external magnetic field. For a single CuPc molecule, the Hamiltonian describing its interaction with the magnetic field is given by $H_{CuPc}$:
\begin{align}
    {\cal H}_{\textrm{CuPc}} =& \beta_e B_0\vec{z} \cdot \mathbf{g} \cdot \vec{S} + \sum_n \vec{S} \cdot \textbf{A}_n \cdot \vec{I}_n+ B_0\vec{z}\cdot \sum_n \gamma_n \vec{I}_n+\sum_n\mathbf{Q}_n\cdot \vec I_{n}+ \sum_{n_i,n_j}\vec{I}_{n_i}\cdot \mathbf{D}_{n_i,n_j}\cdot\vec{I}_{n_j},
    \label{eq:hamilton_cun}
\end{align}
in Eq.~\eqref{eq:hamilton_cun}. This Hamiltonian can be diagonalized into its eigenbasis \(\{\ket{\psi_i}\}\) with energy eigenvalues \(\omega_i\).
The electronic transition strength between two eigenstates  is \(\xi_{ij} = \bra{\psi_j} S_{\perp} \ket{\psi_i}\) at the corresponding transition frequency  \(\omega_{ij} = |\omega_i - \omega_j|\).

\begin{figure}[htbp]
    \centering
    \includegraphics[width=0.7\linewidth]{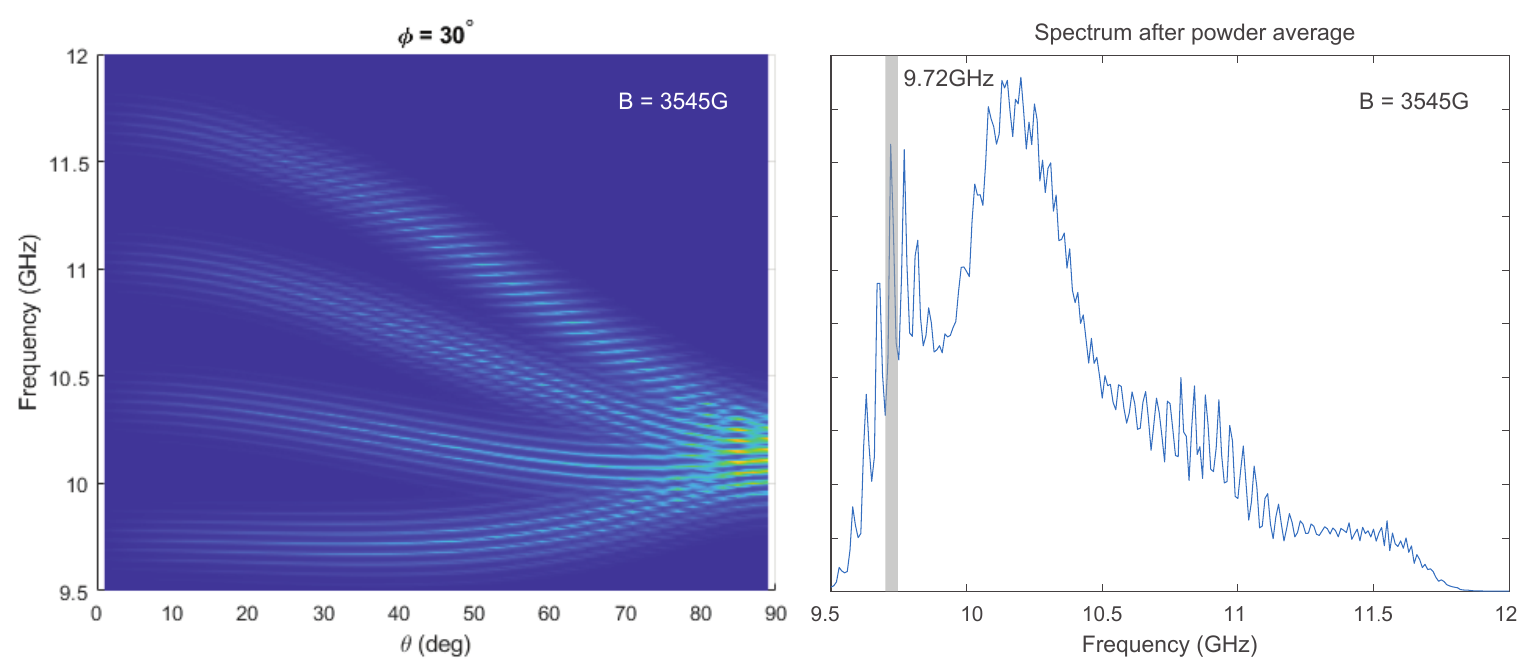}
    \caption{\textbf{Simulation of single CuPc electron spectrum in frequency domain at 3545G.} 
    Left: Example of CuPc spin spectrum at different orientation due to anisotropy. Right: CuPc spectrum after powder average, where the hyperfine manifold at 9.72~GHz is consistent with experiment condition.}
    \label{figS:simu_spec}
\end{figure}

The spin spectral density of the electron spin describes the transition strength distribution in the frequency domain. 
For a single CuPc molecule, the  spectral density \(\xi(\vec B, \omega)\) depends on both the magnitude and orientation of the applied magnetic field.
Due to the strong anisotropy of CuPc and the random orientation of molecules in powder samples, the observed spectrum is obtained by averaging over all possible orientations. Defining \((\theta, \phi)\) the magnetic field direction in the spherical coordinate system of the molecular frame, the powder-averaged spectral density is given by:
\begin{equation}
    S(|\vec B|,\,\omega) = \int_0^{2\pi} d\phi \int_0^{\pi} \sin\theta \, d\theta \, \xi(\vec B,\,\omega).
\end{equation}
\(\xi(\vec B, \omega)\) for each $\theta\in[0^{\circ},90^{\circ}]$ and $\phi\in[0^{\circ},90^{\circ}]$ are recorded for the application  of nuclear spin bath and electron path CCE simulation. Here we plot the spectrum at $B=3545$~G for randomly chosen $\phi = 30^{\circ}$ with varying $\theta$, and spectrum after powder average (Fig.~S\ref{figS:simu_spec}). The shaded area is the hyperfine manifold driven in the experiments at 9.72~GHz.

\section{Spin Echo of a Single Electron Spin in CuPc \label{appendix:spin_echo}}

As mentioned in the main text, we simulate the spin echo signal of a single CuPc molecule to support the conclusion that the strongly hyperfine-coupled nitrogen and copper nuclei contribute only weakly to the transverse decoherence of the electron spin. This simulation is performed from first principles.

The initial state of the spin system in a single CuPc molecule is modeled as
\[
\rho_{\mathrm{ini}} = \frac{1}{2}(\mathbb{I}_e + r \sigma_z) \otimes \frac{\mathbb{I}_{n}}{D},
\]
where $\mathbb{I}_{n}$ is the identity in the $D$-dimensional nuclear spin Hilbert space, $\mathbb{I}_e$ is the  identity operator and $\sigma_z$  the Pauli $z$ operator for the electron spin. The parameter $r$ represents the finite polarization of the electron spin, which we take to be $r = 1$ without loss of generality.

The spin echo sequence consists of a combination of microwave pulses and free evolution intervals. In the experiment, the microwave cavity applies pulses with a magnetic field component perpendicular to the external static field, as described by the Hamiltonian $H_{\mathrm{\mu w}}$ in
\begin{equation}
   {\cal H}_{\mu w} = \Omega_{\mu w} \cos(\omega_{\mu w} t + \phi_{\mu w}) S_{\perp}.
    \label{eq:hamiltonian_pulse}
\end{equation}
The evolution during the microwave pulse can be  computed by going into the rotating frame and taking the rotating wave approximation (RWA), neglecting counter-rotating and non-secular hyperfine terms ($S_{\pm}$) that are averaged out over the short pulse duration.
In the lab frame, the evolution operator for a microwave pulse starting at time \(t_0\) and lasting for a duration \(t_p\) is given by
\begin{equation}
\begin{aligned}
    \hat{U}_p&(t_0, t_p) = \\
    &e^{-i \omega_{\mathrm{\mu w}} S_z (t_0 + t_p)} \cdot e^{-i ({H}_{\parallel} + \Omega_{\mathrm{\mu w}} S_x - \omega_{\mu w}S_z) t_p} \cdot e^{i \omega_{\mathrm{\mu w}} S_z t_0},
\end{aligned}
\end{equation}
where \(\omega_{\mathrm{\mu w}}\) is the microwave driving frequency and \(\Omega_{\mathrm{\mu w}}\) is the Rabi frequency. $H_{\parallel}$ is the  longitudinal component of the Hamiltonian, \([H_{\parallel},S_z]=0\).
We instead retain the full Hamiltonian $H$ during the free evolution.
The pulse durations in our experiment are $t_{\pi/2} = 16$~ns and $t_{\pi} = 32$~ns.

\begin{figure}[htbp]
    \centering
    \includegraphics[width=0.7\linewidth]{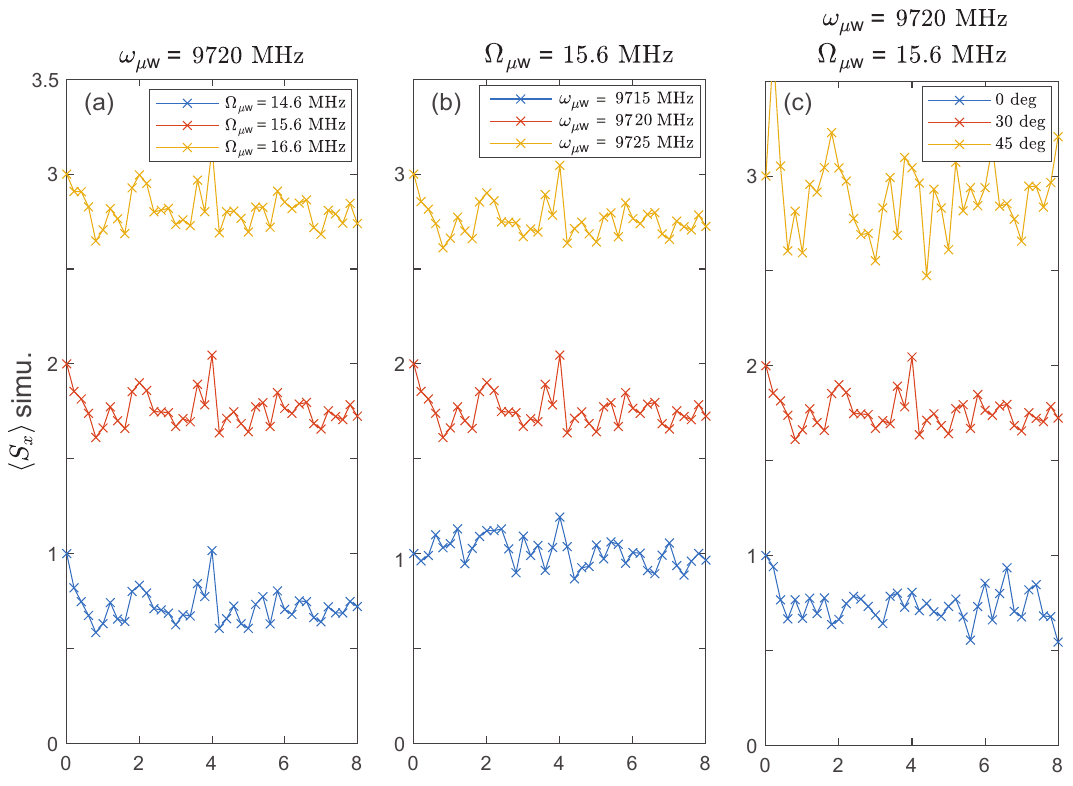}
    \caption{\textbf{Simulation of single CuPc electron spin echo signal.} 
   Simulation result (normalized to $t =0$) at varying microwave power condition (a), varying microwave frequency condition (b), and varying CuPc orientation respective to the external magnetic field. The angle in (c) is the $\phi$ angle of the external magnetic in the molecule frame while $\theta$ is averaged over $[0,90^{\circ}]$. A phenomenological modulation is observed and is highly relates to the CuPc orientation due to the anisotropic hyperfine interactions. }
    \label{figS:simu_echo}
\end{figure}

The total evolution operator for the full spin echo sequence with total free evolution time \(\tau\) is
\begin{align}
   \hat{U}_{\mathrm{echo}}(\tau) = &\hat U_p(t_\pi+t_{\pi/2}+\tau,0)\cdot e^{-i H_0 \frac{\tau}{2}} \cdot \\&\hat{U}_p\left(t_{\pi/2} + {\tau}/{2},\, t_{\pi}\right) \cdot e^{-i H_0 \frac{\tau}{2}} \cdot \hat{U}_p(0,\, t_{\pi/2}).\nonumber
\end{align}
The observable \(\hat{O}_e\) acting only on the electron spin is measured at the end of the spin echo sequence. Its expectation value is given by
\begin{equation}
    \langle \hat{O}_e \rangle(\tau) = \mathrm{Tr}\left[ \hat{O}_e \, \hat{U}_{\mathrm{echo}}(\tau)\, \rho_{\mathrm{ini}}\, \hat{U}_{\mathrm{echo}}^{\dagger}(\tau) \right].
\end{equation}
This result corresponds to a single CuPc molecule at a specific orientation \((\theta, \phi)\) relative to the external static magnetic field.
The powder-averaged expectation value is computed by integrating over all possible molecular orientations:
\begin{equation}
    \langle \hat{O} \rangle_{\mathrm{avg}}(\tau) = \int_0^{2\pi} d\phi \int_0^{\pi} \sin\theta \, d\theta \ \langle \hat{O} \rangle_e(\tau; \theta, \phi).
\end{equation}
The power-averaged result of $\braket{S_x}$ and $\braket{S_z}$is shown in the maintext. In Fig.~S\ref{figS:simu_echo}, we show the simulation result of the $\braket{S_x}$ under varying condition, including imperfect driving condition ($\Omega_{\mu w}$), microwave driving frequency ($\omega_{\mu w}$), and partial CuPc orientation.  

\section{Formulism of Spin Locking in quantum-classical spin bath\label{appendix:spinlocking}}

In the main text, we decomposed the Hamiltonian into a set of 2-dimensional Hilbert subspaces--effective electron spin-\(1/2\) systems with a defined nuclear spin states. This effective electron spin serves as the central spin and couples to a surrounding spin bath as described by the central spin-spin bath model. In the rotating frame defined by the driving frequency \(\omega_{\mathrm{\mu w}}\), the total Hamiltonian is given by 
\begin{align}
{\cal H}_n &={\cal H}_0+ {\cal H}_{e\text{-}n,q} + {\cal H}_{e\text{-}n,c}, \label{eq:hamilton_nuclearspin}
\end{align}
\begin{align}
{\cal H}_{e\text{-}n,c} &= \gamma_e \tilde{B}_z(t) S_z\label{eq:hamiltonian_nc},
\end{align}
\begin{align}
{\cal H}_{e\text{-}n,q} &= \gamma_H{B}_0\sum_{n}I^n_z+ S_z \sum_{n}\sum_{\sigma = x,y,z} A_{z\sigma}^{n} I_{\sigma}^{n} , \label{eq:hamiltonian_nq}
\end{align}
and the microwave driving field applied during the spin-locking pulse is described by
\begin{equation}
    {\cal H}_{\text{locking}} = \Omega_1 S_x.\label{eq:hamiltonian_locking_p}
\end{equation}
The distinct short- and long-time behavior of the spin-locking signal enables separate analysis of the contributions from classical and quantum fluctuations of the spin bath.

\subsection{Classical Spin Bath Effect on Spin-Locking Signal}

The classical spin bath acts as a quasi-static noise source that leads to depolarization of the central spin in the transverse plane during spin locking. In the rotating frame, the Hamiltonian of the driven central spin yields two dressed states, \(\ket{e}\) and \(\ket{g}\), separated by an energy gap \(\Omega_r = \sqrt{\Omega_1^2 + \delta^2}\), where \(\delta\) is the detuning frequency as defined in the main text.

In the dressed-state basis, the effective Hamiltonian describing the central spin subject to a fluctuating longitudinal magnetic field \(\tilde{B}(t)\) (Eq.~\eqref{eq:hamiltonian_nc}) can be written as:
\begin{equation}
\begin{aligned}
    {\cal H} =& \frac{1}{2}\hbar\left(\Omega_r + \frac{\delta}{\Omega_r}\tilde{B}(t)\right)\left(\ketbra{e}{e} - \ketbra{g}{g}\right) \\
    &+ \frac{1}{2}\hbar \frac{\Omega_1}{\Omega_r} \gamma_e \tilde{B}(t) \left(\ketbra{e}{g} + \ketbra{g}{e}\right).
\end{aligned}
\end{equation}

The second term induces incoherent transitions between the dressed states, leading to spin relaxation in the rotating frame. Using Fermi's golden rule, the longitudinal relaxation rate under spin locking is given by:
\begin{equation}
\begin{aligned}
    \frac{1}{T_{1\rho}} &= \Gamma_{e \rightarrow g} + \Gamma_{g \rightarrow e} \\
    &= \frac{1}{\hbar^2} \int_{-\infty}^{+\infty} dt\, e^{i \Omega_r t} 
    \left\langle H_{eg}(0) H_{eg}^\dagger(t) + H_{ge}(0) H_{ge}^\dagger(t) \right\rangle \\
    &= \frac{1}{2} \left( \frac{\Omega_1}{\Omega_r} \right)^2 \gamma_e^2 
    \int_{-\infty}^{+\infty} dt\, e^{i \Omega_r t} \langle \tilde{B}(t) \tilde{B}(0) \rangle \\
    &= \frac{1}{2} \left( \frac{\Omega_1}{\Omega_r} \right)^2 \gamma_e^2 \int_{-\infty}^{+\infty} dt\, e^{i \Omega_r t} G(t) \\
    &= \frac{1}{2} \left( \frac{\Omega_1}{\Omega_r} \right)^2 \gamma_e^2 S(\Omega_r),
\end{aligned}
\label{eq:fermi}
\end{equation}
where \(G(t) = \langle \tilde{B}(t) \tilde{B}(0) \rangle\) is the autocorrelation function of the classical noise field, and \(S(\Omega_r)\) is the corresponding power spectral density evaluated at the dressed-state splitting frequency \(\Omega_r\).

\subsection{Quantum Spin Bath Effect on Spin-Locking Signal}

For a central spin strongly coupled to a finite number of nuclear spins forming a spin cluster, the dynamics are governed by the Hamiltonian \(H_0 + H_{e\text{-}n,q} + H_{\text{locking}}\), as defined in the main text [Eqs.~\eqref{eq:hamilton_nuclearspin}, \eqref{eq:hamiltonian_nq}, and \eqref{eq:hamiltonian_locking_p}]. 

To illustrate the quantum effect of the spin bath on spin locking, we consider the simplest case where the bath consists of a single nuclear spin. In the rotating frame, the total Hamiltonian becomes
\begin{equation}
{\cal H} = \delta S_z + \Omega_1 S_x + S_z (A_{zz} I_z + A_{zx} I_x + A_{zy} I_y) + \omega_H I_z,
\label{eq:hamiltonian_nq1H}
\end{equation}
where \(\delta\) is the detuning of the central spin, \(\Omega_1\) is the Rabi frequency, \(A_{\mu\nu}\) are the hyperfine coupling constants, and \(\omega_H\) is the nuclear Larmor frequency. We define the complex combination \(A_{z\pm} = A_{zx} \pm i A_{zy}\).

Here we define a tilted rotating frame, which is a rotation about the \(y\)-axis by angle \(\theta_{\pm}\):
\begin{equation}
\sin\theta_{\pm} = \frac{\Omega_1}{\omega_{\mathrm{eff}}^{\pm}} = \frac{\Omega_1}{\sqrt{(\delta \pm A_{zz})^2 + \Omega_1^2}},
\end{equation}
where
\begin{equation}
\label{eq-6-7}
\omega_{\mathrm{eff}}^{\pm} = \sqrt{(\delta \pm A_{zz})^2 + 2 \Omega_1^2}.
\end{equation}

In this basis, the Hamiltonian (Eq.~\eqref{eq:hamiltonian_nq1H}) can be written in a block-matrix form as
\begin{equation}
\label{eq:hamiltonian_nq1Ht}
\tfrac{\hbar}{2} \left(
\begin{matrix}
\omega_{\mathrm{eff}}^+ + \omega_H & 0 & \tfrac{1}{2} A_{z+} \cos\theta & -\tfrac{1}{2} A_{z+} \sin\theta \\
0 & -\omega_{\mathrm{eff}}^+ + \omega_H & -\tfrac{1}{2} A_{z-} \sin\theta & -\tfrac{1}{2} A_{z+} \cos\theta \\
\tfrac{1}{2} A_{z-} \cos\theta & -\tfrac{1}{2} A_{z+} \sin\theta & \omega_{\mathrm{eff}}^- - \omega_H & 0 \\
-\tfrac{1}{2} A_{z-} \sin\theta & -\tfrac{1}{2} A_{z-} \cos\theta & 0 & -\omega_{\mathrm{eff}}^- - \omega_H
\end{matrix}
\right),
\end{equation}
where \(\theta = \tfrac{1}{2}(\theta_+ + \theta_-)\).

We assume the initial state of the system is a partially polarized central spin along the \(x\)-axis, represented by the density matrix \(\frac{1}{2}(\mathbb{I} + r \sigma_x)\), while the nuclear spin is unpolarized. In the tilted frame, the full initial density matrix becomes
\begin{equation}
\label{eq:densitymatrix_nq1Ht}
\rho_0^t =
\left(
\begin{array}{cccc}
\frac{1 + r \cos\theta_+}{4} & -\frac{r \sin\theta_+}{4} & 0 & 0 \\
\frac{r \sin\theta_+}{4} & \frac{1 - r \cos\theta_+}{4} & 0 & 0 \\
0 & 0 & \frac{1 + r \cos\theta_-}{4} & -\frac{r \sin\theta_-}{4} \\
0 & 0 & \frac{r \sin\theta_-}{4} & \frac{1 - r \cos\theta_-}{4}
\end{array}
\right).
\end{equation}

We now focus on the central \(2\times2\)  subspace of Eq.~\eqref{eq:hamiltonian_nq1Ht} (zero-quantum, ZQ), which governs flip-flop dynamics between the dressed states. This sub-Hamiltonian can be expressed as an effective two-level system (fictitious spin \(s\)):
\begin{equation}
\begin{aligned}
\mathcal{H}_s = \tfrac{\hbar}{2} \left(
\begin{matrix}
\omega_{\mathrm{eff}}^+ - \omega_H & -\tfrac{1}{2} A_{z-} \sin\theta \\
-\tfrac{1}{2} A_{z+} \sin\theta & -\omega_{\mathrm{eff}}^- + \omega_H
\end{matrix}
\right)\\
\\
= \hbar \left[ \tfrac{\Delta \omega}{2} \sigma_z - \tfrac{1}{4} |A_{z-}| \sin\theta \, \sigma_x + \tfrac{\Delta \omega_{\mathrm{eff}}}{2} \mathbb{I} \right],
\end{aligned}
\end{equation}
where \(\Delta \omega = \omega_{\mathrm{eff}} - \omega_H\), \(\omega_{\mathrm{eff}} = \tfrac{1}{2}(\omega_{\mathrm{eff}}^+ + \omega_{\mathrm{eff}}^-)\), and \(\Delta \omega_{\mathrm{eff}} = \tfrac{1}{2}(\omega_{\mathrm{eff}}^+ - \omega_{\mathrm{eff}}^-)\). Here, \(\sigma_x\) and \(\sigma_z\) are Pauli operators, and \(\mathbb{I}\) is the identity operator.

According to Eq.~\eqref{eq:densitymatrix_nq1Ht}, the initial state this fictitious spin is $\tilde{\rho}(0)=\tfrac{1}{2}(\mathbb{I}+\tilde{r}\sigma_z)$, where $\tilde{r} = r \cdot \frac{\cos\theta_+ + \cos\theta_-}{2}$ is the initial polarization.
The polarization dynamics under the subspace evolves as
\begin{equation}
\begin{aligned}
    P_s(t) = \text{Tr}[\tfrac{1}{2}\sigma_z e^{-\frac{i}{\hbar}H_st}\tilde{\rho}(0)e^{-\frac{i}{\hbar}H_st}]\\
    = \tilde{r} \left[1 - \frac{|A_{z-}|^2}{4\alpha_-^2}(1 - \cos \alpha_- t)\right],
\end{aligned}
\end{equation}
where the oscillation frequency \(\alpha_-\) is given by
\begin{equation}
\alpha_- = \sqrt{(\omega_{\mathrm{eff}} - \omega_H)^2 + \tfrac{1}{4} |A_{z-}|^2 \sin^2\theta}.
\end{equation}

Similarly, the double-quantum (DQ) subspace, associated with simultaneous flip-flip transitions, contributes an opposite polarization component, with characteristic frequency
\begin{equation}
\alpha_+ = \sqrt{(\omega_{\mathrm{eff}} + \omega_H)^2 + \tfrac{1}{4} |A_{z-}|^2 \sin^2\theta}.
\end{equation}

The total polarization transfer from both ZQ and DQ processes is therefore
\begin{equation}
\label{equ:Ps}
P_s(t) = \tilde{r} \left[ \frac{|A_{z-}|^2}{4\alpha_+^2}(1 - \cos \alpha_+ t) - \frac{|A_{z-}|^2}{4\alpha_-^2}(1 - \cos \alpha_- t) \right].
\end{equation}
For X-band EPR, where \(|A_{z-}|^2 \ll \alpha_+^2\), the double-quantum contribution is negligible, resulting in a single-frequency oscillation dominated by the zero-quantum transition. This behavior is consistent with our numerical simulations (see Supplementary Materials).

For quantum clusters involving multiple nuclear spins, an analytical expression becomes intractable. However, as discussed in the main text, the distribution of hyperfine coupling strengths leads to a superposition of multiple single-frequency oscillations. This superposition results in an overall decay of the spin polarization, which is also confirmed by our numerical simulations.

\section{Spin Locking Experiment and Simulation}
\subsection{Experimental Data}
\begin{figure}[htbp]
    \centering
    \includegraphics[width=0.7\linewidth]{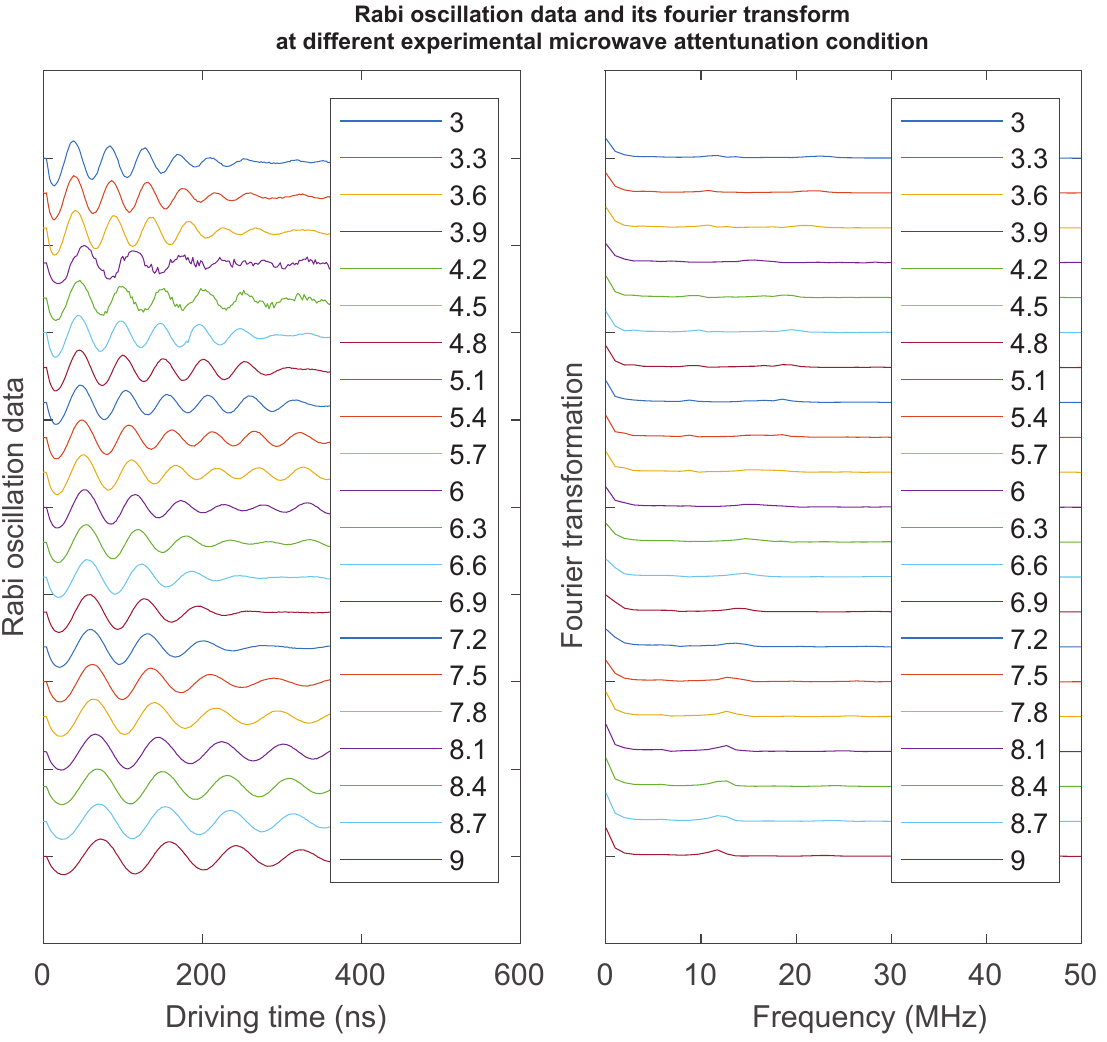}
    \caption{\textbf{Rabi oscillation calibration.} 
    Rabi oscillation data and corresponding Fourier transforms at different driving powers. Each curve corresponds to a different microwave power attenuation. Larger attenuation values correspond to lower microwave power. The Rabi frequencies extracted from Fourier analysis are used for simulations and subsequent data analysis.}
    \label{figS:spin_locking_rabi}
\end{figure}

Spin-locking experiments were conducted at 5~K. The microwave driving strength for each sample was pre-calibrated using Rabi oscillation measurements, with representative data shown in Fig.~S\ref{figS:spin_locking_rabi}.  

\begin{figure}[htbp]
    \centering
    \includegraphics[width=0.7\linewidth]{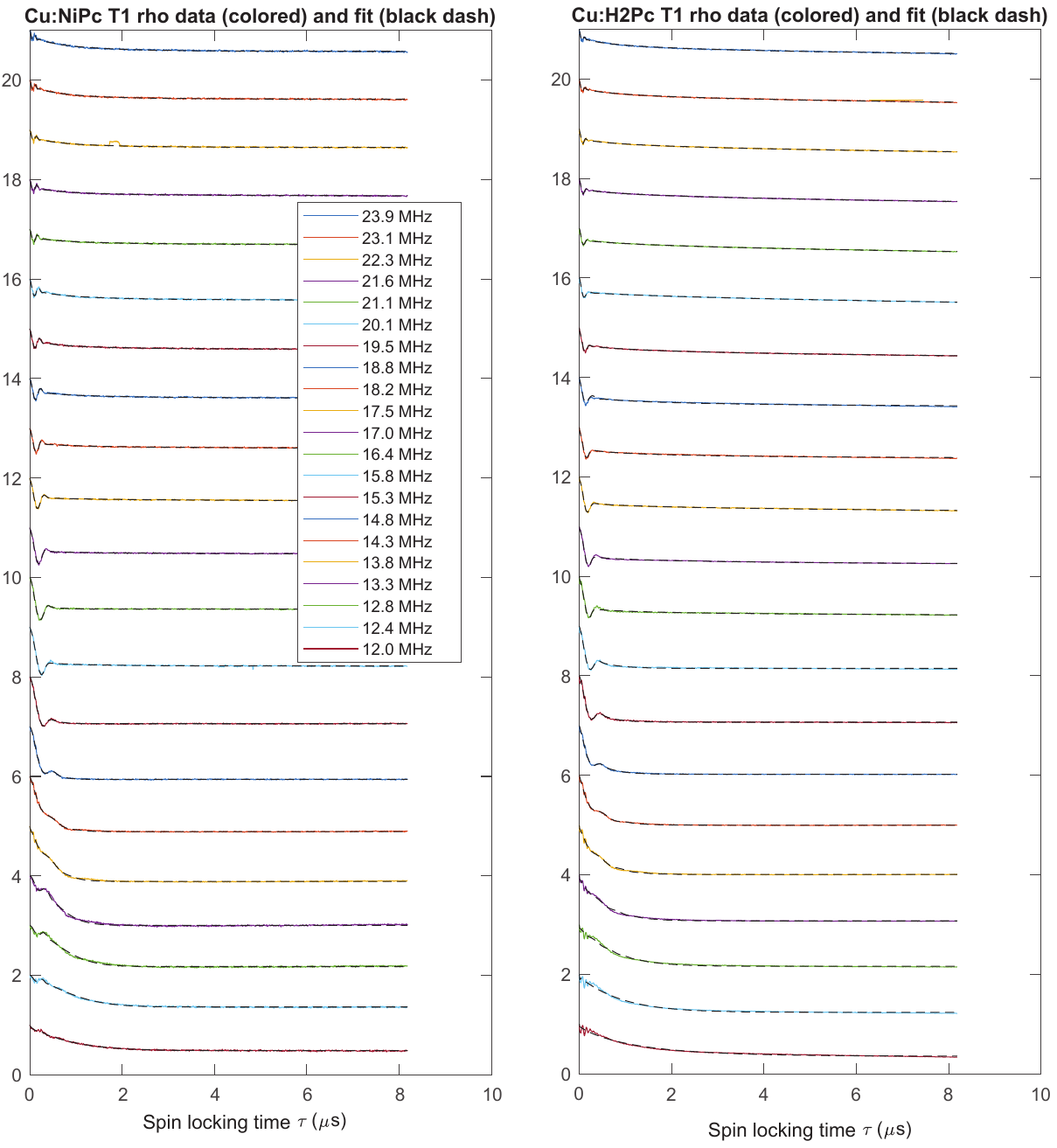}
    \caption{\textbf{Spin-locking measurements.} Experimental spin-locking results for CuPc:NiPc (left) and CuPc:H$_2$Pc (right) at different locking pulse amplitudes. Solid lines are fits to Eq.~(number to be added) in the main text.}
    \label{figS:spin_locking_exp}
\end{figure}

In the spin-locking sequence, the initial $\pi/2$ pulse (duration 16~ns) prepares the electron spin into transverse magnetization. The evolution of this transverse magnetization as a function of locking pulse duration, for different driving strengths, is presented in Fig.~S\ref{figS:spin_locking_exp} for CuPc:NiPc and CuPc:H$_2$Pc matrices. The data were fitted using Eq.~(a) in the main text.  

\begin{table*}[b]
\renewcommand\arraystretch{1.4}
\caption{The distance ($d$) and hyperfine coupling strength ($A = \sqrt{A_{\parallel}^2 + A_{\perp}^2}$) between the electron spin of CuPc and nearby nuclear spins in the $\beta-$ phase H$_2$Pc (NiPc) crystal lattice.}
\centering
\begin{tabular}{ cccc  }
 \hline
 \hline
 \multirow{1}{6cm}{\centering Nuclear spin}  & \multirow{1}{2cm}{\centering Quantity} & \multirow{1}{2cm}{\centering $d/\mathring{A}$} & \multirow{1}{4cm}{\centering $A/kHz$}\\
 \hline
 \multirow{1}{6cm}{\centering N from CuPc not bounded with Cu}  & \multirow{1}{2cm}{\centering 4} & \multirow{1}{2cm}{\centering 3.38 (3.39)} & \multirow{1}{4cm}{\centering 294.46 (289.21)}\\
 \hline
 \multirow{1}{6cm}{\centering H from CuPc at near end}  & \multirow{1}{2cm}{\centering 8} & \multirow{1}{2cm}{\centering 5.91 (5.90)} & \multirow{1}{4cm}{\centering 765.19 (770.42)}\\
 \hline
 \multirow{1}{6cm}{\centering H from CuPc at far end}  & \multirow{1}{2cm}{\centering 8} & \multirow{1}{2cm}{\centering 7.60 (7.57)} & \multirow{1}{4cm}{\centering 359.61 (363.70)}\\
 \hline
 \multirow{1}{6cm}{\centering H from H$_2$Pc at the center}  & \multirow{1}{2cm}{\centering 4} & \multirow{1}{2cm}{\centering 4.71 (5.33)} & \multirow{1}{4cm}{\centering 1513.36 (1039)}\\
 \hline
 \multirow{1}{6cm}{\centering N from XPc bounded with Cu}  & \multirow{1}{2cm}{\centering 4} & \multirow{1}{2cm}{\centering 4.08 (4.05)} & \multirow{1}{4cm}{\centering 167.73 (172.35)}\\
 \hline
 \multirow{1}{6cm}{\centering N from XPc not bounded with Cu}  & \multirow{1}{2cm}{\centering 2} & \multirow{1}{2cm}{\centering 3.30 (3.22)} & \multirow{1}{4cm}{\centering 319.30 (340.92)}\\
 \hline
 \multirow{1}{6cm}{\centering H from XPc at near end}  & \multirow{1}{2cm}{\centering 4} & \multirow{1}{2cm}{\centering 4.44 (4.44)} & \multirow{1}{4cm}{\centering 1811.62 (1799.39)}\\
 \hline
 \multirow{1}{6cm}{\centering H from XPc at far end}  & \multirow{1}{2cm}{\centering 4} & \multirow{1}{2cm}{\centering 6.19 (6.26)} & \multirow{1}{4cm}{\centering 665.25 (644.03)}\\
 \hline
 \hline
\end{tabular}
\label{tables:Hyperfine}
\end{table*}

\subsection{Simulation of the Quantum Spin Bath}
As described in the main text, the quantum spin bath is modeled by including the nearest-neighbor hydrogen nuclei surrounding the CuPc electron spin in the CuPc:XPc lattice.  

\begin{figure}[htbp]
    \centering
    \includegraphics[width=0.7\linewidth]{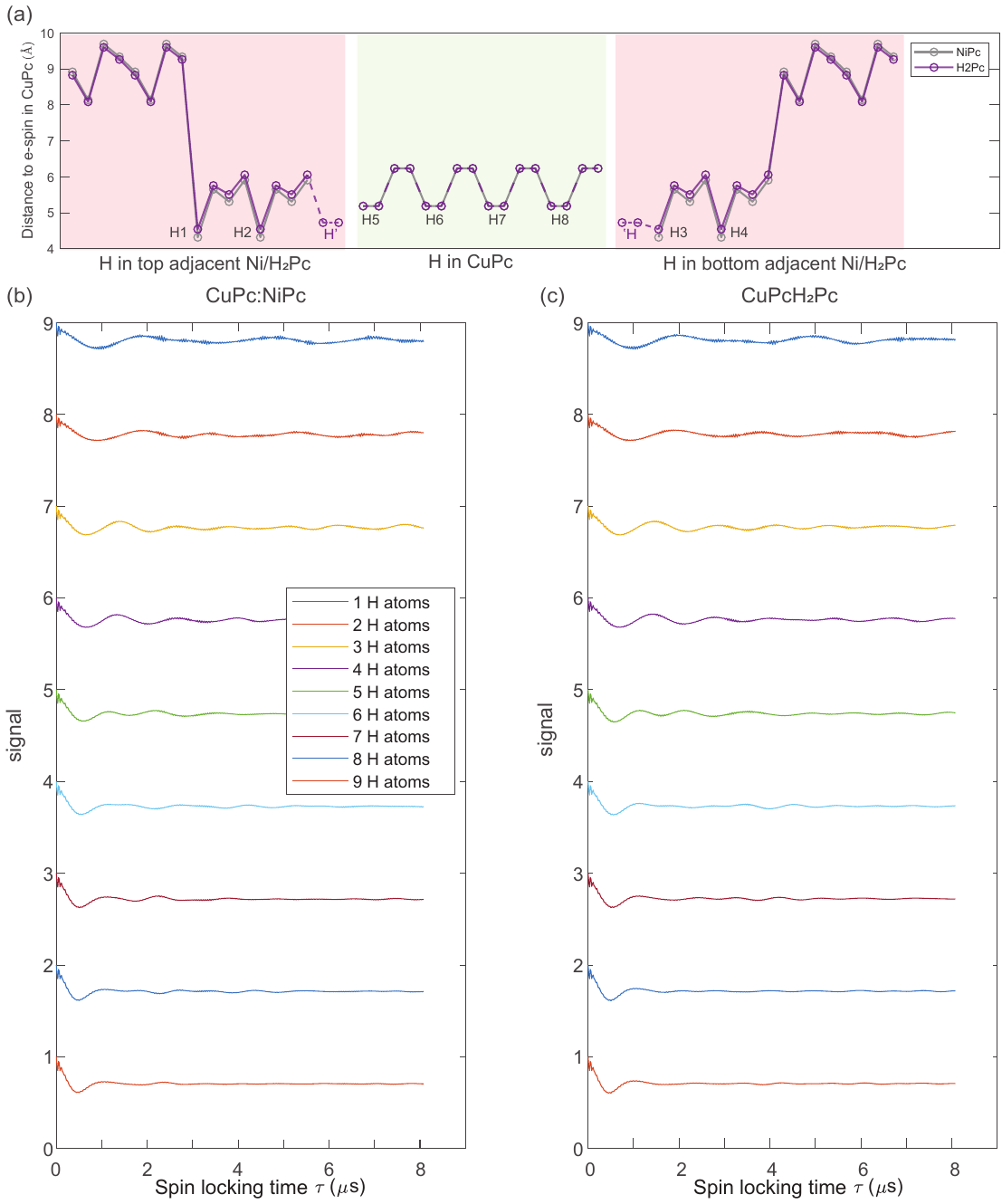}
    \caption{\textbf{Spin-locking simulations with varying numbers of hydrogen spins in the quantum spin cluster.} 
    (a) Distances between CuPc electron spins and surrounding hydrogens from CuPc and neighboring NiPc/H$_2$Pc molecules. The closest hydrogens are grouped as H1-H8, with an additional H$'$ group present only in H$_2$Pc, corresponding to central hydrogens. All other groups correspond to edge hydrogens of Pc molecules.  
    (b, c) Simulated spin-locking dynamics for CuPc:NiPc (b) and CuPc:H$_2$Pc (c) at resonance condition $\Omega_1 = 15.09$~MHz. The number of included hydrogens is increased sequentially according to their distance from the electron spin.}
    \label{figS:spin_locking_sim_H_num}
\end{figure}

The insets of Figure~S\ref{figS:cce_hydrogen} shows the relative positions and hyperfine couplings of these hydrogens, including atoms from both the CuPc molecule and its two nearest-neighbor XPc molecules. Based on interaction strength, the hydrogens are labeled H1-H16. For both NiPc and H$_2$Pc matrices, the closest hydrogens are consistently located at the molecular edges of the adjacent Pc molecules. The distances are shown in Figure~S\ref{figS:spin_locking_sim_H_num}(a) 

We first analyze the effect of cluster size by increasing the number of included nuclear spins. As an example, simulations were performed at a locking pulse strength of $\Omega_1 = 14.7$~MHz. When only the strongest-coupled hydrogen (H1) is included, the simulation agrees with the analytical solution presented in the Appendix of the main text. Adding more nuclear spins sequentially, ordered by their distance from the electron spin, produces damped oscillations. The results converge when the cluster size exceeds eight spins, indicating that larger clusters do not significantly alter the dynamics.  

Figure~S\ref{figS:spin_locking_sim_H_num}(a) summarizes the relevant distances. For CuPc:NiPc, the spin bath consists of 10 dominant hydrogens, with the eight strongest hyperfine couplings being most significant: four from the neighboring NiPc molecule (H1-H4) and four from CuPc itself (H5-H8). For groups H5-H8, each site corresponds to two possible hydrogens; in simulations we included two from H5 and H7, and one from H6 and H8. For CuPc:H$_2$Pc, the spin bath contains four edge hydrogens (H1-H4), two groups of central hydrogens from adjacent H$_2$Pc molecules, and two hydrogens from CuPc itself.  

Next, we included detuning and ensemble averaging. For a given molecular orientation relative to the external magnetic field, the electron-spin detuning was obtained from simulated CuPc spectra, with resonance frequency $\Delta \in 9790 \pm 15.6$~MHz. The $\pm 15.6$~MHz range corresponds to power broadening from the $\pi/2$ pulse. For each orientation, the spin-locking signal was averaged over the corresponding detuning values, and the final simulated signal was obtained by averaging over all molecular orientations.  

Figures~S\ref{figS:spin_locking_sim_H_num}(b, c) present the simulation results for NiPc and H$_2$Pc, respectively. As the number of hydrogens increases, the oscillations become more strongly damped, and long-time coherence is suppressed—consistent with the experimental observations. Convergence is again reached for cluster sizes larger than eight spins.  

\begin{figure}[htbp]
    \centering
    \includegraphics[width=0.7\linewidth]{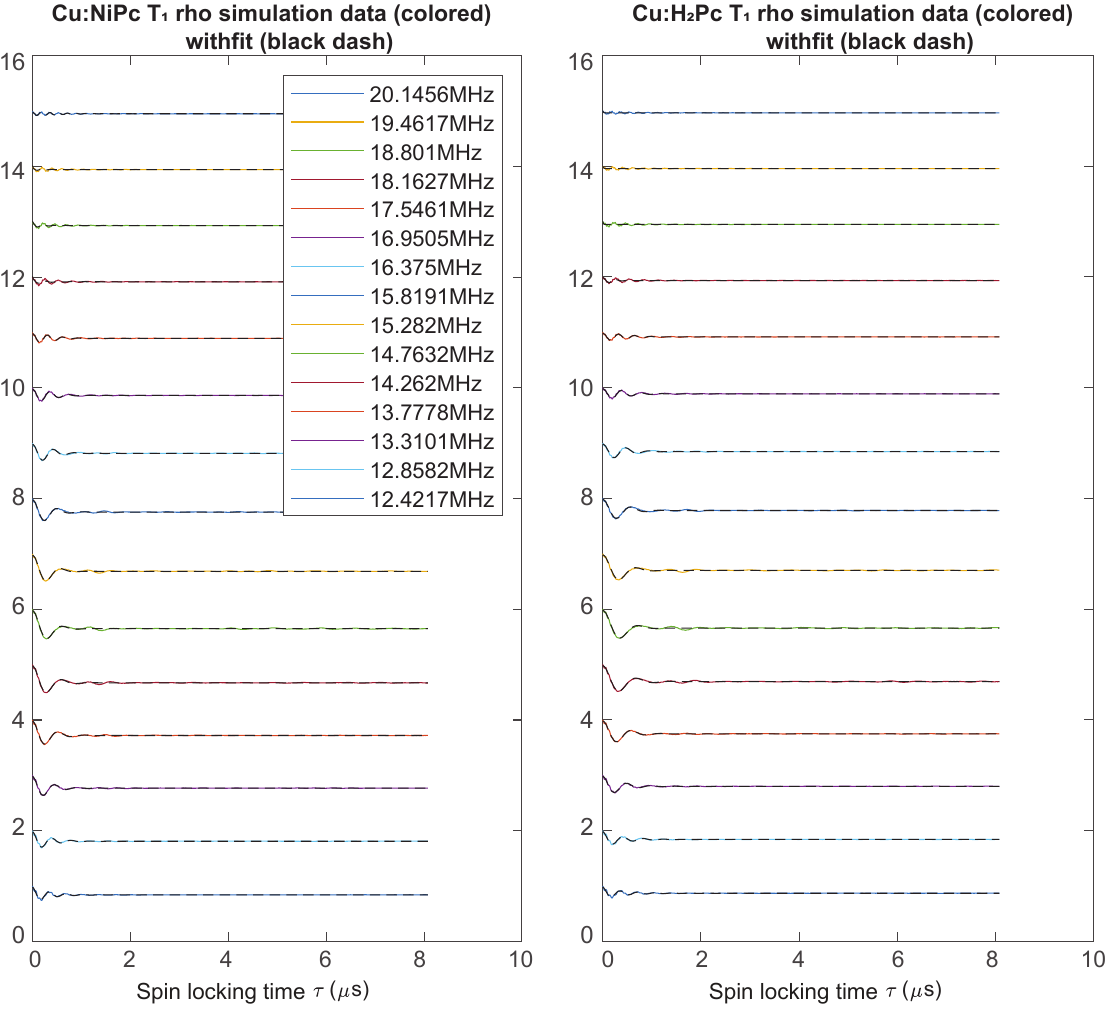}
    \caption{\textbf{Spin-locking simulations including ensemble averaging.} 
    Simulated spin-locking results for CuPc:NiPc (left) and CuPc:H$_2$Pc (right) at different locking pulse amplitudes. Orientation averaging over all molecular configurations is included. The simulated data are fitted using Eq.~(number to be added) in the main text.}
    \label{figS:spin_locking_sim_final}
\end{figure}

Finally, we simulated the full spin-locking dynamics as a function of driving strength, applying the same fitting procedure as in the experiments. The results are summarized in Fig.~S\ref{figS:spin_locking_sim_final} together with the corresponding fits.

\section{Formulism of Effective Electron Spin Bath}

In the main text, we model the system as comprising a central effective spin-\(1/2\) (i.e., the EPR transition that is resonantly driven in the spin echo sequence, labeled \(\iota_R\)) and a bath consisting of effective spin-\(1/2\) transitions from all other CuPc molecules. The nuclear spins are effectively frozen during the electron spin evolution. Then the spin operator of each bath CuPc molecule can be represented as a direct sum over all \(M\) two-level subspaces: 
\[
\vec{S}^k = \bigoplus_{\iota=1}^{M} \vec{S}^{k,\iota},
\]
where \(k\) indexes different CuPc molecules and \(\vec{S}^{k,\iota}\) denotes the spin-\(1/2\) operator for the \(\iota\)th EPR transition (two-level subspace) on the $k$th CuPc molecule. Correspondingly, the density matrix of the CuPc spin bath is given by
\begin{equation}
    \rho_b = \bigotimes_{k=1}^{N} \left[ \frac{1}{M^N} \bigoplus_{\iota=1}^{M} \rho_{k,\iota} \right] = \bigotimes_{k=1}^{N} \rho_k,
\end{equation}
where \(N\) is the total number of CuPc molecules, and \(\rho_{k,\iota}\) is a \(2 \times 2\) density matrix representing the \(\iota\)th effective two-level subsystem of the \(k\)th CuPc. The full density matrix of molecule \(k\), \(\rho_k\), is the direct sum over its \(M\) subspaces. The prefactor \(1/M^N\) assumes uniform initial occupation of all the energy levels.

The system Hamiltonian, including dipolar interactions among CuPc electron spins, is then given by:
\begin{equation}
\begin{aligned}
   H_e =\ & \Delta_{\iota_R} S_z^{\iota_R} + \sum_k \vec{S}^{\iota_R} \cdot \mathbf{D}^k \cdot \bigoplus_{\iota} \vec{S}^{k,\iota} \\
   & + \sum_k \bigoplus_{\iota} \left( \Delta_{\iota} S_z^{k,\iota} + \sum_{k'} \bigoplus_{\iota'} \vec{S}^{k',\iota'} \cdot \mathbf{D}^{k'k} \cdot \vec{S}^{k,\iota} \right),
\end{aligned}
\label{eq:hamilton_electronspin}
\end{equation}
where \(\mathbf{D}^k\) is the dipolar coupling tensor between the central spin and the spin at site \(k\), and \(\mathbf{D}^{k'k}\) represents the dipolar interaction between bath spins on different CuPc molecules.
In the interaction picture, the Hamiltonian becomes
\begin{equation}
\begin{aligned}
   \tilde{H}_e = &\sum_k \tilde{\vec{S}}^{\iota_R} \cdot \mathbf{D}^k \cdot \bigoplus_{\iota} \tilde{\vec{S}}^{k,\iota} \\
   &+ \sum_k \sum_{k'} \bigoplus_{\iota} \tilde{\vec{S}}^{k,\iota} \cdot \mathbf{D}^{kk'} \cdot \bigoplus_{\iota'} \tilde{\vec{S}}^{k',\iota'},
\end{aligned}
\end{equation}
where
\begin{equation}
\begin{aligned}
    \tilde{\vec{S}}^{\iota_R} &= e^{i \Delta_{\iota_R} S_z^{\iota_R} t} \vec{S}^{\iota_R} e^{-i \Delta_{\iota_R} S_z^{\iota_R} t},\\
\tilde{\vec{S}}^{k,\iota} &= e^{i \Delta_{\iota} S_z^{k,\iota} t} \vec{S}^{k,\iota} e^{-i \Delta_{\iota} S_z^{k,\iota} t}.
\end{aligned}
\end{equation}

At any time \(t\), the total density matrix in the interaction picture can be written as a convex sum over product states of the central spin (c) and the spin bath (b):
\[
\tilde{\rho}(t) = \sum_w p_w\, \tilde{\rho}_c^w(t) \otimes \tilde{\rho}_b^w(t),
\]
where \(p_w \geq 0\), \(\sum_w p_w = 1\), and \(w\) indexes the decomposition terms.

For each \(w\), the bath density matrix can be expanded as

\begin{equation}
\begin{aligned}
\tilde{\rho}_c^w(t) &\otimes \tilde{\rho}_b^w(t) = \tilde{\rho}_c^w(t) \bigotimes_{k=1}^{N} \left[ \frac{1}{M^N} \bigoplus_{\iota=1}^{M} \tilde{\rho}_{k,\iota}^w \right]\\
&= \frac{1}{M^N} \tilde{\rho}_c^w(t) \bigotimes \left[ \bigoplus_{\{\vec{\iota}\}} \bigotimes_{k=1}^{N} \tilde{\rho}_{k,\vec{\iota}_k}^w \right].
\end{aligned}
\end{equation}

Here \(\{\vec{\iota}\}\) denotes the set of all possible index combinations \((\iota_1, \iota_2, \ldots, \iota_N)\), where each \(\iota_k \in \{1, \ldots, M\}\) corresponds to one of the \(M\) hyperfine-split EPR transitions assigned to the \(k\)th CuPc molecule.

\begin{widetext}
The Hamiltonian terms can be similarly decomposed. For the central spin-bath interaction:

    \begin{equation}
    \begin{aligned}
        \tilde{H}_1 = \sum_k\sum_{\mu,\nu\in\{x,y,z\}} {D}^k_{\mu,\nu}
   \tilde{{S}}^{\iota_R}_\mu  \bigotimes\left[\bigoplus_{\iota}  \tilde{{S}}^{k,\iota}_{\nu}\right]
   =\sum_k {D}^k_{\mu,\nu}\sum_{\mu,\nu}
   \tilde{{S}}^{\iota_R}_\mu  \bigotimes\left\{ \bigoplus_{\{\iota\}}
        \left[\bigotimes_{\kappa=1}^{k-1}  \mathbb{I}\,\bigotimes  \tilde{S}^{k,\vec{\iota}_\kappa}_\nu \bigotimes_{\kappa=k+1}^{N}  \mathbb{I}\right]\right\} ,
    \end{aligned}
\end{equation}
and for bath-bath interactions:
\begin{equation}
    \begin{aligned}
        \tilde{H}_2&= \sum_k \sum_{k'>k} \sum_{\mu,\nu\in\{x,y,z\}}{D}^{kk'}_{\mu\nu}\left[\bigoplus_{\iota}  \tilde{\vec{S}}^{k,\iota}_\mu\right]\bigotimes\left[\bigoplus_{\iota'}\tilde{\vec{S}}^{k',\iota'}\right]\\
         &=\sum_k \sum_{k'>k} \sum_{\mu,\nu}{D}^{kk'}_{\mu\nu}\left\{ \bigoplus_{\{\vec{\iota}\}}
        \left[\bigotimes_{\kappa=1}^{k-1}  \mathbb{I}\,\bigotimes  \tilde{S}^{k,\vec{\iota}_{\kappa}}_\nu \bigotimes_{\kappa=k+1}^{N}  \mathbb{I}\right]\right\}\bigotimes\left\{ \bigoplus_{\{\vec{\iota}\}}
        \left[\bigotimes_{\kappa=1}^{k-1}  \mathbb{I}\,\bigotimes  \tilde{S}^{k',\vec{\iota}_{\kappa'}}_\nu \bigotimes_{\kappa'=k'+1}^{N}  \mathbb{I}\right]\right\}\\
        &=\sum_k \sum_{k'>k}\sum_{\mu,\nu} {D}^{kk'}_{\mu\nu}\left\{ \bigoplus_{\{\vec{\iota}\}}
        \left[\bigotimes_{\kappa=1}^{k-1}  \mathbb{I}\,\bigotimes  \tilde{S}^{k,\vec{\iota}_{\kappa}}_\nu \bigotimes_{\kappa=k+1}^{k'-1}  \mathbb{I}\,\bigotimes  \tilde{S}^{k',\vec{\iota}_{\kappa}}_\nu \bigotimes_{\kappa=k'+1}^{N}  \mathbb{I}\right]\right\}\\
    \end{aligned}
\end{equation}
\end{widetext}

The reduced density matrix of the central spin is given by
\[
\tilde{\rho}_c(t) = \mathrm{Tr}_b[\tilde{\rho}(t)] = \sum_w p_w \tilde{\rho}_c^w(t),
\label{eq:density_matrix_app_e}
\]
and its evolution is governed by
\[
\dot{\tilde{\rho}}_c(t) = -i\, \mathrm{Tr}_b\left\{ [\tilde{H}_e, \tilde{\rho}(t)] \right\}.
\]

For each \(w\) term in Eq.~\eqref{eq:density_matrix_app_e}, we can obtain:
\begin{widetext}
    \begin{equation}
\begin{aligned}
    \dot{\tilde{\rho}}&_c^w(t) = -i\, \mathrm{Tr}_b\left\{[\tilde{H}_e,\, \tilde{\rho}_c^w(t) \otimes \tilde{\rho}_b^w(t)]\right\}\\
    &=-\frac{i}{M^N}\sum_{\{\vec{\iota}\}}\left\{\sum_k\sum_{\mu,\nu}\text{Tr}_b\left[\tilde{\rho}_c^w(t)\otimes\tilde{\rho}^{k,\iota_k},  {D}^k_{\mu,\nu}
   \tilde{{S}}^{\iota_R}_\mu  \otimes  \tilde{{S}}^{k,\iota_k}_{\nu}\right]+\tilde{\rho}_c^w(t)\sum_k\sum_{k'>k}\sum_{\mu,\nu}\text{Tr}\left[\tilde{\rho}^{k,\iota_k}(t)\otimes\tilde{\rho}^{k',\iota_{k'}},  {D}^{k,k'}_{\mu,\nu}
  \tilde{{S}}^{k,\iota_k}_\mu  \otimes  \tilde{{S}}^{k',\iota_{k'}}_{\nu}\right]\right\}\\
   &=-\frac{i}{M^N}\sum_{\{\vec{\iota}\}}\left\{\text{Tr}_b\left[\tilde{\rho}_c^w(t)\bigotimes_{k=1}^N\tilde{\rho}^{k,\iota_k},  \sum_k\sum_{\mu,\nu}{D}^k_{\mu,\nu}
   \tilde{{S}}^{\iota_R}_\mu  \otimes  \tilde{{S}}^{k,\iota}_{\nu}\right]+\tilde{\rho}_c^w(t)\text{Tr}\left[\bigotimes_{k=1}^N\tilde{\rho}^{k,\iota_k},  \sum_k\sum_{k'>k}\sum_{\mu,\nu}{D}^{k,k'}_{\mu,\nu}
  \tilde{{S}}^{k,\iota_k}_\mu  \otimes  \tilde{{S}}^{k',\iota_{k'}}_{\nu}\right]\right\}
\end{aligned}
\end{equation}
\end{widetext}

It is clear that, aside from the summation over \(\{\vec{\iota}\}\) and the prefactor \(1/M^N\), the remaining terms are identical to those obtained by modeling each CuPc molecule as occupying a single spin transition subspace. By definition, \(\vec{\iota}\) corresponds to a specific spin configuration of occupied EPR transitions across the ensemble, and there are \(M^N\) distinct microscopic configurations in total. Thus, the averaging over \(\{\vec{\iota}\}\) explicitly captures the ensemble averaging over all possible spin state occupations at the CuPc sites. The validity of the effective spin-1/2 model extends to systems comprising two types of CuPc orientations, each exhibiting distinct sets of transition frequencies due to anisotropy.

In our CCE simulations, we begin by randomly sampling CuPc occupation sites within the XPc crystal structure. Since in $\beta$-phase CuPc:XPc crystal, the lattice contains two types of oriented CuPc molecules. For CuPc molecules located on the up-tilted columns (as illustrated in Figure~1), we assign a transition frequency randomly drawn from the simulated spectrum corresponding to this specific orientation. For the other type of CuPc orientation, we similarly assign transition frequencies from its own distinct spectrum shifted by the hyperfine couplings. 

The spin echo dynamics are then simulated for each specific spin configuration. To capture ensemble averaging, we repeat the simulation over many instances by resampling both the spin transition frequencies and the spatial positions of the bath spins.

We provide an intuitive physical interpretation. At any given moment, the spin bath can only be at a specific spin configuration: each CuPc molecule is in one of its hyperfine-shifted EPR transition subspace. Since all \(M\) transitions are equally probable, the initial spin configuration of the CuPc ensemble corresponds to a mixture of \(M\) distinct spin-\(1/2\) species, each with a different resonance frequency and a fractional density . We have shown that different transitions on the same CuPc evolve independently, so each molecule evolution remains in subspace corresponding to its initial state throughout the evolution.
Under spatial ensemble averaging, repeated experimental realizations, or thermal fluctuations over time, all microscopic configurations are collectively sampled, reflecting the statistical nature of the fully mixed state. This behavior is conceptually analogous to statistical polarization, in which fluctuations in the net magnetization of a fully mixed spin ensemble arise from the definite spin orientations present in each microscopic configuration.

\section{CCE Simulation Details for the Hydrogen Spin Bath}
The CCE method is a well-established approach for simulating the spin-echo evolution of an electron spin interacting with a nuclear spin bath. In this work, we employed second-order CCE (CCE-2) to simulate the CuPc electron spin dynamics in hydrogen nuclear spin baths for both H$_2$Pc and NiPc matrices.

\begin{figure}[htbp]
    \centering
    \includegraphics[width=0.7\linewidth]{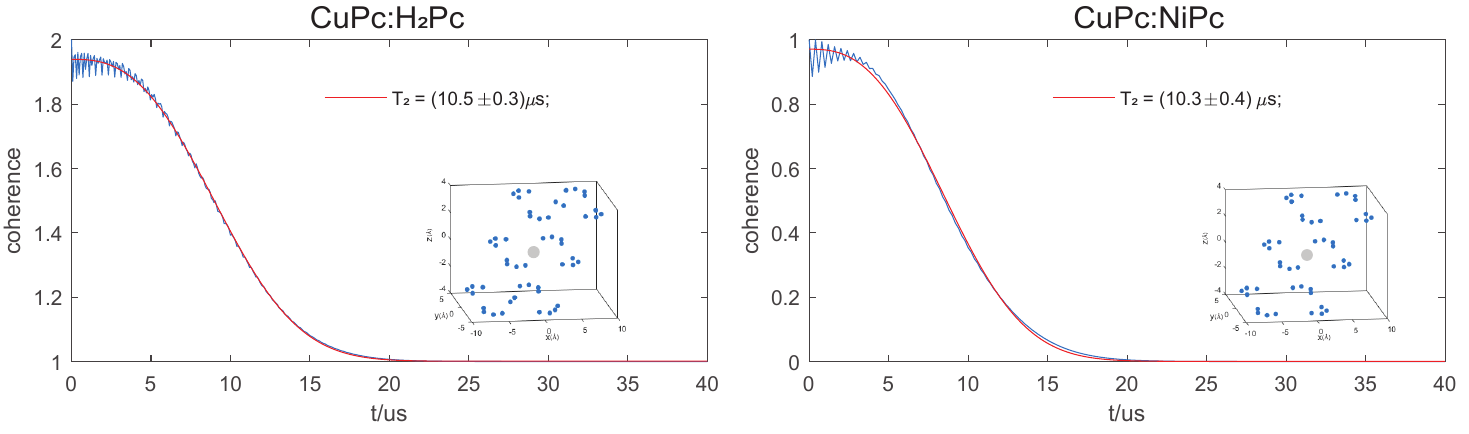}
    \caption{\textbf{CCE simulation of the hydrogen nuclear spin bath.} 
    Inset: repeat unit cell showing the spatial distribution of hydrogen nuclei (blue dots) around the CuPc electron spin (gray dot).}
    \label{figS:cce_hydrogen}
\end{figure}

The maximum bath size was set to $r_{\mathrm{bath}} = 40$~\AA, and the maximum nuclear-nuclear dipolar interaction distance was $r_{\mathrm{dipole}} = 6$~\AA. The spatial distribution of hydrogen atoms and the corresponding simulation results are shown in Fig.~S\ref{figS:cce_hydrogen}.  

\section{CCE Simulation Details for the Electron Spin Bath}
\begin{figure}[htbp]
    \centering
    \includegraphics[width=0.7\linewidth]{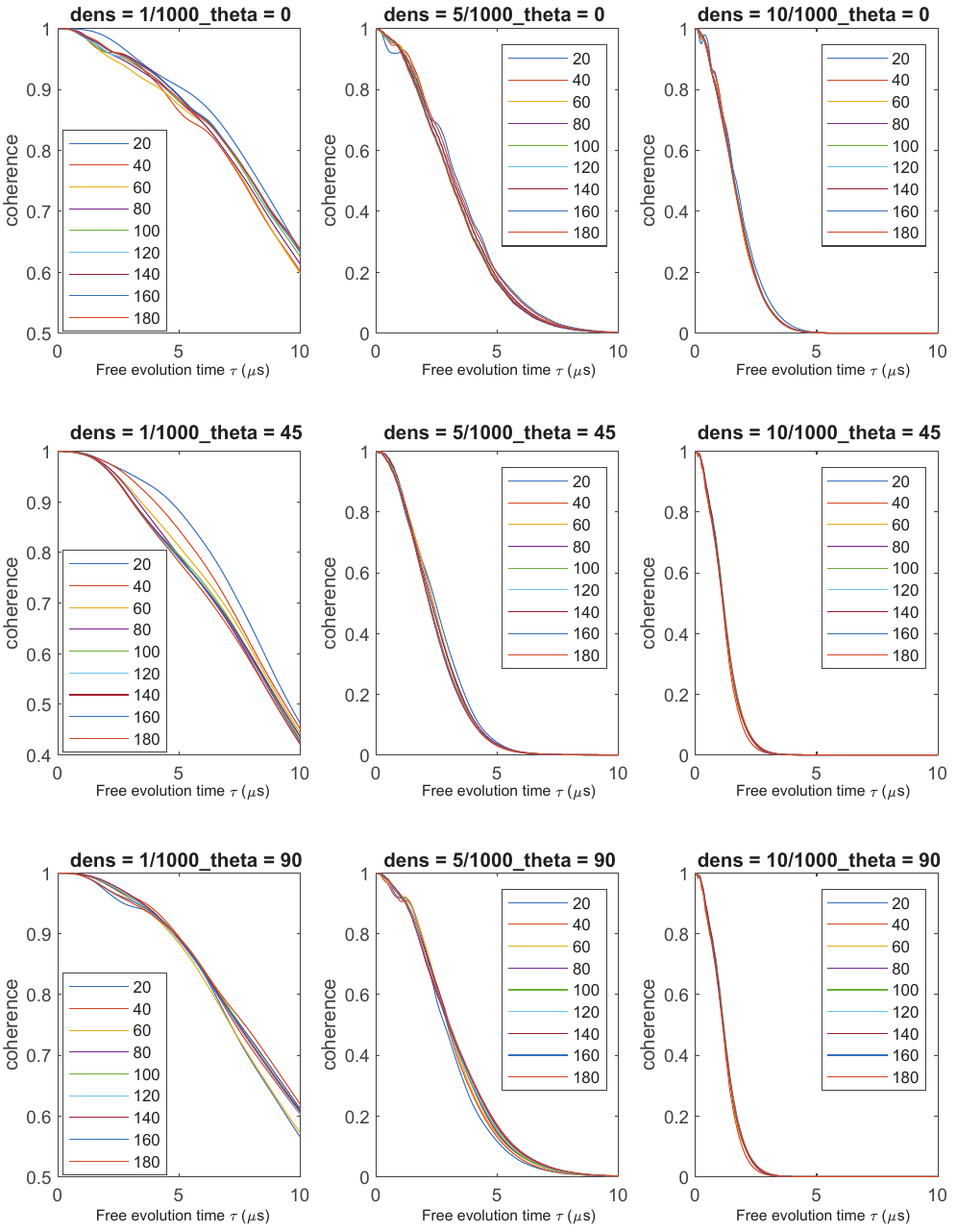}
    \caption{\textbf{CCE convergence with respect to ensemble averaging.} 
    For different CuPc dilution ratios and crystal orientations, the results converge once the number of random spin-bath samplings exceeds $\sim$180.}
    \label{figS:cce_electron_sampling_conv}
\end{figure}

To simulate the electron spin bath using the CCE method, we first examined the convergence of ensemble averaging with respect to the number of sampling realizations.  
Figure~S\ref{figS:cce_electron_sampling_conv} shows results for three CuPc dilution ratios at different CuPc crystal orientations. Convergence is reached when the number of random spin-bath samplings exceeds $\sim$120 in all cases.  

\begin{figure}[htbp]
    \centering
    \includegraphics[width=0.7\linewidth]{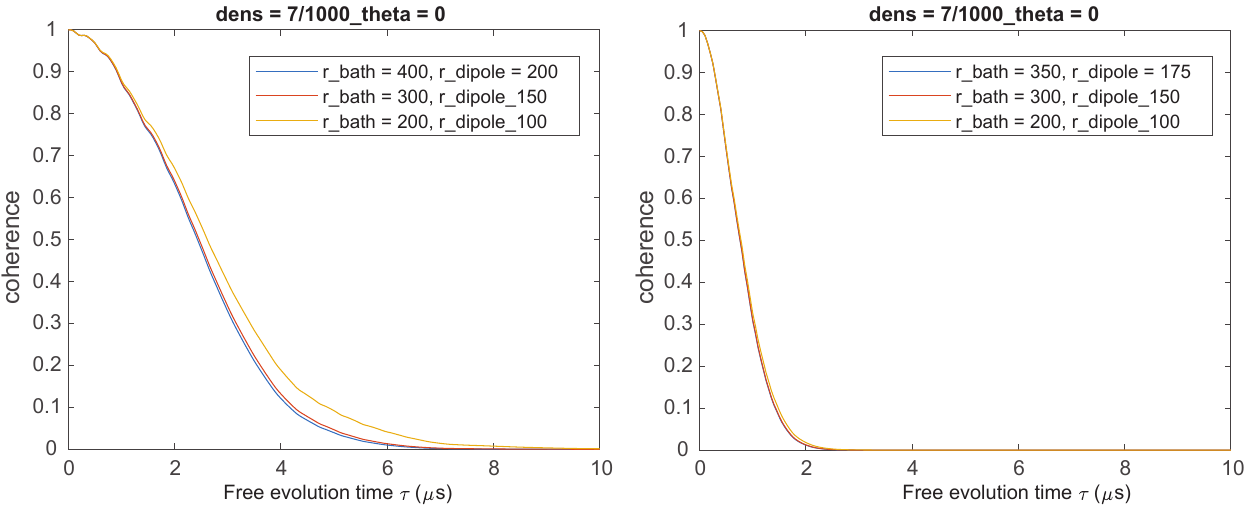}
    \caption{\textbf{CCE convergence with respect to bath size and dipolar cutoff.} 
    Results for low (7/1000) and high (20/1000) CuPc dilution ratios at different $r_{\mathrm{bath}}$ and $r_{\mathrm{dipole}}$ combinations. Convergence is reached under the conditions specified in the text.}
    \label{figS:cce_electron_rbrd_conv}
\end{figure}

Next, we tested convergence with respect to the bath size ($r_{\mathrm{bath}}$) and the maximum dipolar cutoff distance ($r_{\mathrm{dipole}}$). For low (7/1000) and high (20/1000) CuPc dilution ratios, simulations were performed for three combinations of $r_{\mathrm{bath}}$ and $r_{\mathrm{dipole}}$, as shown in Fig.~S\ref{figS:cce_electron_rbrd_conv}. These results provide the convergence benchmarks used for all dilution ratios in the main text. Specifically, the parameters used were:  
\[
\begin{aligned}
&2/1000 \;\rightarrow\; (r_{\mathrm{bath}} = 400, \; r_{\mathrm{dipole}} = 200) \\
&7/1000 \;\rightarrow\; (r_{\mathrm{bath}} = 300, \; r_{\mathrm{dipole}} = 150) \\
&13/1000 \;\rightarrow\; (r_{\mathrm{bath}} = 300, \; r_{\mathrm{dipole}} = 150) \\
&20/1000 \;\rightarrow\; (r_{\mathrm{bath}} = 250, \; r_{\mathrm{dipole}} = 125) \\
&25/1000 \;\rightarrow\; (r_{\mathrm{bath}} = 250, \; r_{\mathrm{dipole}} = 125) \\
&28/1000 \;\rightarrow\; (r_{\mathrm{bath}} = 200, \; r_{\mathrm{dipole}} = 100)
\end{aligned}
\]

\begin{figure}[htbp]
    \centering
    \includegraphics[width=0.7\linewidth]{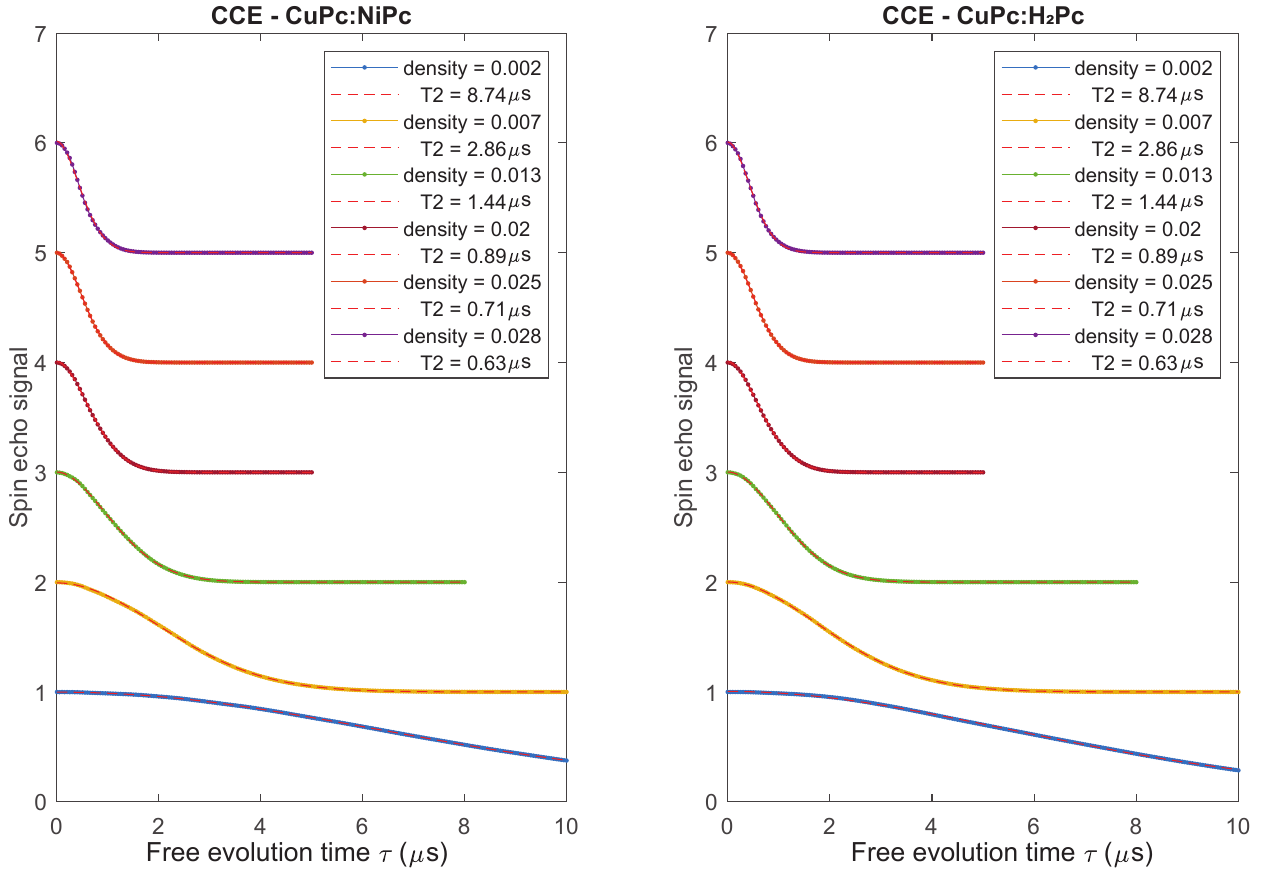}
    \caption{\textbf{CCE simulation of electron spin bath dynamics.} 
    Final CCE simulations with fits for different matrices and CuPc dilution ratios.}
    \label{figS:cce_electron_final}
\end{figure}

The final CCE simulation results under these convergence conditions are presented in Fig.~S\ref{figS:cce_electron_final}.  

\bibliography{CuPc_EPR_ref}